\documentclass[pdflatex,sn-mathphys-num]{sn-jnl}

\usepackage{graphicx}%
\usepackage{multirow}%
\usepackage{amsmath,amssymb,amsfonts}%
\usepackage{amsthm}%
\usepackage{mathrsfs}%
\usepackage[title]{appendix}%
\usepackage{xcolor}%
\usepackage{textcomp}%
\usepackage{manyfoot}%
\usepackage{booktabs}%
\usepackage{algorithm}%
\usepackage{algorithmicx}%
\usepackage{algpseudocode}%
\usepackage{listings}%
\usepackage[normalem]{ulem}

\usepackage{subcaption} 

\usepackage[version=3]{mhchem} 
\usepackage{graphicx}
\usepackage{dcolumn}
\usepackage{bm}
\usepackage{siunitx}
\usepackage{amsmath}
\usepackage[dvipsnames]{xcolor}
\usepackage[version=3]{mhchem} 
\usepackage{siunitx}  

\usepackage{comment}
\usepackage{multirow}
\usepackage{array}
\usepackage{textpos}
\usepackage{rotating} 
\usepackage{esvect}

\usepackage{graphicx}
\usepackage{dcolumn}
\usepackage{bm}

\usepackage[utf8]{inputenc}
\usepackage{float}
\usepackage[T1]{fontenc}
\usepackage{mathptmx}
\usepackage[noabbrev, capitalise]{cleveref}
\usepackage{amssymb}
\usepackage{natbib}

\theoremstyle{thmstyleone}%
\theoremstyle{thmstyletwo}%

\theoremstyle{thmstylethree}

\renewcommand{\tablename}{Extended Table}

\begin{document}

\title[Article Title]{On-chip nanoplasma for adaptive electromagnetic protection}

\author[1]{\fnm{} \sur{Ruiqi Huang}} 
\equalcont{These authors contributed equally to this work.}

\author[1]{\fnm{} \sur{Hanqing Liu}}
\equalcont{These authors contributed equally to this work.}

\author[2]{\fnm{} \sur{Kui Dang}}
\equalcont{These authors contributed equally to this work.}

\author[3,4]{\fnm{} \sur{Yue Lang}}
\author[3,4]{\fnm{} \sur{Ruiyang Peng}}
\author[5]{\fnm{} \sur{Qihui Zhou}}
\author[1]{\fnm{} \sur{Qian Dong}}

\author[1]{\fnm{} \sur{Xiaocheng Ni}}

\author[1]{\fnm{} \sur{Yuan Xu}}

\author*[1]{\fnm{} \sur{Jibin Liu}}
\email{liujibin@nudt.edu.cn}

\author[1]{\fnm{} \sur{Yanlin Xu}}

\author[1]{\fnm{} \sur{Chenxi Liu}}

\author*[1]{\fnm{} \sur{Song Zha}}
\email{zhasong0551@163.com}

\author*[2]{\fnm{} \sur{Jincheng Zhang}}
\email{jchzhang@xidian.edu.cn}

\author*[1]{\fnm{} \sur{Peiguo Liu}}
\email{pg731@126.com}

\affil[1]{\orgdiv{College of Electronic Science and Technology}, \orgname{National University of Defense Technology}, \orgaddress{\city{Changsha}, \postcode{410073}, \state{Hunan}, \country{China}}}

\affil[2]{\orgdiv{State Key Laboratory of Wide-Bandgap Semiconductor Devices and Integrated Technology}, \orgname{Xidian University}, \orgaddress{\city{Xi'an}, \postcode{710071}, \state{Shannxi}, \country{China}}}

\affil[3]{\orgdiv{College of Science}, \orgname{National University of Defense Technology}, \orgaddress{\city{Changsha}, \postcode{410073}, \state{Hunan}, \country{China}}}

\affil[4]{\orgdiv{Hunan Key Laboratory of Extreme Physics and Applications}, \orgname{National University of Defense Technology}, \orgaddress{\city{Changsha}, \postcode{410073}, \state{Hunan}, \country{China}}}

\affil[5]{\orgdiv{Defense Innovation Institute}, \orgname{Academy of Military Sciences}, \orgaddress{\city{Beijing}, \postcode{100071}, \country{China}}}




\abstract{
Over the past decade, semiconductor diodes have served as the primary switching elements in adaptive electromagnetic (EM) protection\cite{Eleftheriades-1,WakatsuchiKim-338,YangC}, yet their performance has been compromised by parasitic effects and thermal accumulation, rendering them inadequate against the rapidly evolving landscape of high-power microwave (HPM) threats \cite{LiuLiu-2,2024HighPowerMicrowaves}. Here we show that on-chip nanoplasma switches (NPMS), composed of gallium nitride electrodes on silicon carbide substrates, exhibit superior radio frequency (RF) and thermal characteristics, positioning them as ideal field-driven switches in RF front-end protectors. By integrating NPMS into metasurfaces, antennas and circuit limiters, we achieve an adaptive response that ensures low-loss transmission for normal signals and high shielding against HPMs, while offering extended operating bandwidth and substantially higher tolerance than conventional solid‑state devices. This robust, nanoscale structure has significant potential for protecting unmanned aerial vehicles, radars, satellites and other highly integrated platforms requiring strength and stability in EM environments. The findings of this study open up new routes to support EM safety of high-precision detection and imaging for next‑generation RF front ends, with straightforward scalability to millimetre‑wave and terahertz frequencies.

}

\maketitle

\begin{figure}[H]
	\centering
	\includegraphics[width=1\linewidth,angle=0]{"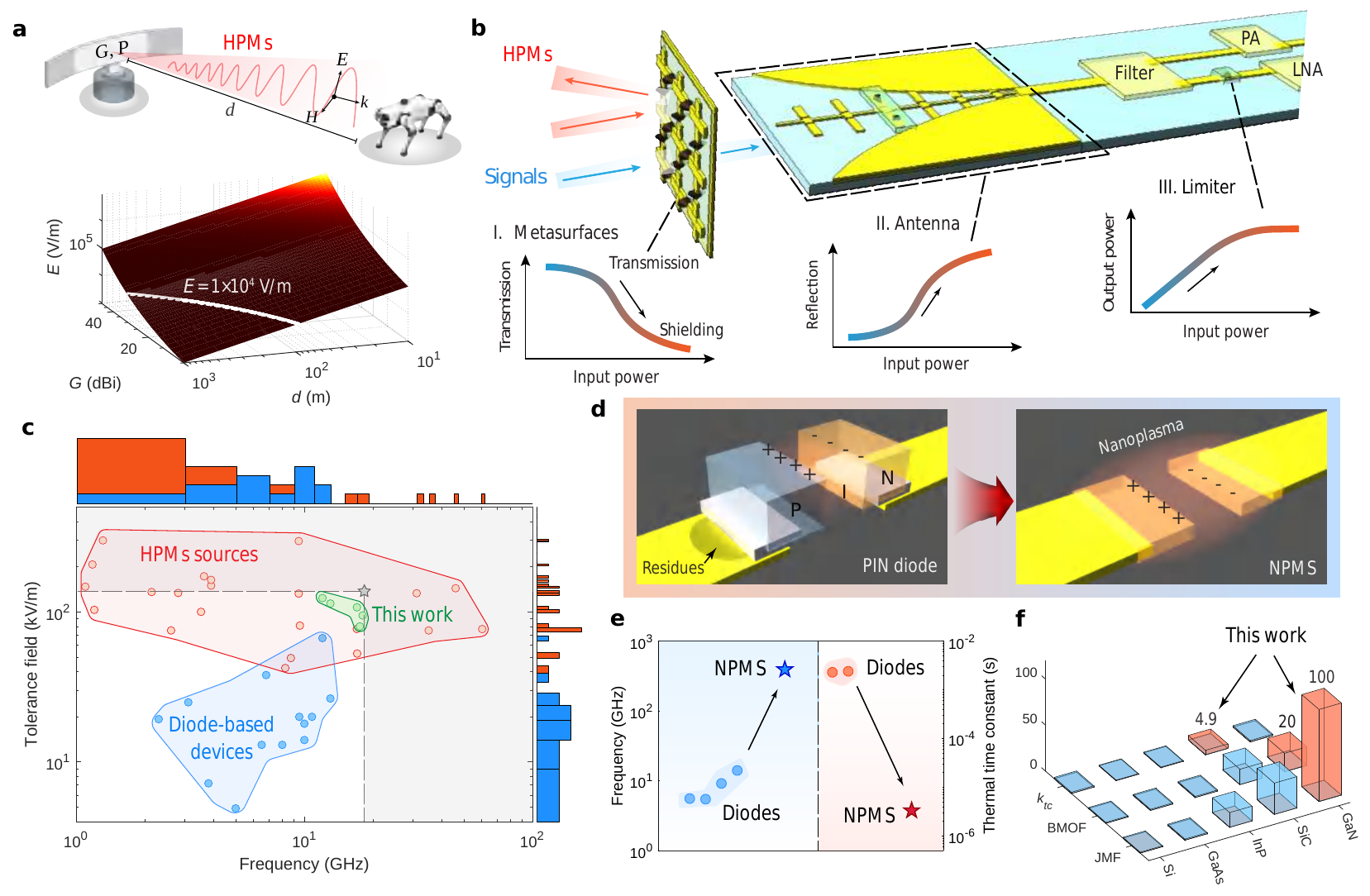"}
	\caption{ \textbf{Adaptive RF front-end protection against HPM threats}. \textbf{a,} Conceptual illustration depicting HPM effects on embodied intelligent agent robot (top) and generated electrical field $E$ on the objective as the function of source gain $G$ and distance $d$ (bottom). \textbf{b,} Main pathways for HPMs suppression in RF front ends: I, metasurface (field); II, antenna (field-circuit); III, limiter (circuit). \textbf{c,} Performance of current diode-based protectors (blue) and our proposed devices (green), compared with fields radiated by HPM sources with 30~\si{dBi} gain at a distance of 100~\si{m} (red); detailed data are provided in Extended Tables; gray region, testing limits; histograms of operating frequency and tolerance field (in log scale) are plotted in side panels. \textbf{d,} Diagrams of welded PIN diode and the on-chip NPMS. \textbf{e,} Calculated efficient switching frequency (left) and thermal time constant (right) for diodes and NPMS. \textbf{f,} Comparison of thermal conductivity ($k_{\text{tc}}$), Johnson Merit Figure (JMF) and Baliga Figure of Merit (BFOM) among semiconducting materials commonly used for RF electronics. References: Si, GaAs, InP, SiC and GaN.          
     }
	\label{fig:1}
\end{figure}

Radio‑frequency (RF) front ends comprise components such as antennas, power amplifiers and low‑noise amplifiers that enable electronic equipment to interface with electromagnetic (EM) fields. Driven by advances in semiconductors and integration technologies, RF front‑end modules are evolving towards higher operating frequencies, increased integration densities and enhanced power‑handling capabilities\cite{PazosFontana-354,VukovicKulmer-355}. These trends, however, increase their susceptibility to external EM interference, particularly high‑power microwaves (HPMs). With nanosecond (ns) rise times and gigawatt (GW) peak amplitudes, HPMs can couple into RF channels and induce irreversible damage in transceiver circuits\cite{LiYuan-326,Rick,ZhenFeng-312}. As illustrated in Fig.~\ref{fig:1}a, an unprotected commercial embodied intelligent robot, for example, ceases operation due to LIDAR and power supply damage upon HPM irradiation at 10$^4$~\si{V/m} (Supplementary Note 1). Consequently, information platforms such as communication, navigation, sensing and detection equipment in close proximity to high‑gain sources are facing an acute need for efficient EM protection.

Such protection adopted in RF front ends requires in-band adaptive ability that allows low-loss transmission of normal signals while instantaneously suppressing the incoming HPMs \cite{LiuLiu-2}. This is considerably more challenging than conventional EM shielding, which simply blocks all incident waves\cite{Xinfeng,Geosan}. Since the early 2010s, nonlinear‑impedance switches have been widely adopted in adaptive EM protectors, including nonlinear metasurfaces\cite{Mengkun,WuLiu-368,ZhangJihong}, reconfigurable antennas\cite{Zhasong,ZhaLiu-374} and circuit limiters\cite{Rodion,ZhaoKang-4,ZhuWang-15}. Semiconductor diodes, whose conductivity is modulated by the incident field, have become the primary choice for such switching elements. As shown in Fig.~\ref{fig:1}b, diode‑based devices can distinguish HPMs from normal signals by input power level, thereby controlling the coupling paths and preventing excessive power injection. However, with the state-of-the-art HPM sources now capable of peak amplitudes exceeding 10~\si{GW} at frequencies up to 140~\si{GHz} \cite{2024HighPowerMicrowaves}, existing approaches still fall short of practical requirements in both operating band and power tolerance (Fig.~\ref{fig:1}c). Parasitic effects from element packaging and soldering degrade switching above 14~\si{GHz}, while contact thermal resistances impede transient heat dissipation to the substrate, reducing HPM tolerance to below 67~\si{kV/m} (Supplementary Note 2).  

Nanoplasma switch (NPMS) offers picosecond response times over a wide dynamic range of power levels, exceeding those of conventional solid-state devices\cite{SamizadehNikooJafari-48}. Recent studies have incorporated NPMS into applications including millimeter-wave and terahertz sources, active-modulation metadevices and gas conversion chips\cite{RezaeiMatioli-45,Guangyu,Samizadeh,Sun}. Their field-emission characteristics and high integration prompt us to explore their potential as a diode replacement in high-performance protection (Fig.~\ref{fig:1}d). Theoretical evaluation indicates that NPMS offer an efficient switching frequency and heat dissipation that are approximately $10^3$ times of magnitude superior to those of commercial diodes (Fig.~\ref{fig:1}e and Supplementary Note 2). However, their use in adaptive protectors remains unexplored and challenging, as NPMS must sustain ultra-fast switching and sufficient tolerance against HPM-induced thermal and electrical stresses. Our initial tests on all-metal NPMS reveal that thermal accumulation from electrode melting tends to induce local short circuits, resulting in severe switching instability (Supplementary Note 3). To address this, wide-bandgap semiconductors offer a promising route to NPMS construction, attributed to their excellent thermal properties, radiation hardness and electron saturation drift velocity (Fig.~\ref{fig:1}f). We thus adopt gallium nitride (GaN) and silicon carbide (SiC) as the emitter and substrate materials of NPMS, respectively. Moreover, upon heating, the strong covalent bonds in these materials undergo bond breaking, which thermodynamically favors decomposition over melting under ambient pressure\cite{Bouazizi,Wataru}, thereby eliminating melt-induced shorting and enhancing device reliability.

Here we introduce NPMS as an alternative to semiconductor diodes used as field-driven impedance-tunable components in adaptive EM protectors. Our design encompasses the main coupling pathways in RF front end, including nonlinear metasurfaces (spatial field), reconfigurable antennas (field-circuit conversion) and circuit limiters. Despite limitations imposed by the measurement setup, the proposed protectors exhibit markedly enhanced performance compared with diode-based devices, including operation across the Ku band, ultra-high tolerance (Fig.~\ref{fig:1}c) and reliable operation under repetitive pulsed excitation. These results establish on-chip nanoplasma as a practical and scalable route towards adaptive protection for high-integration electronic systems, with implications for EM security of future information platforms.
\\

\begin{figure}
	\centering
	\includegraphics[width=1\linewidth,angle=0]{"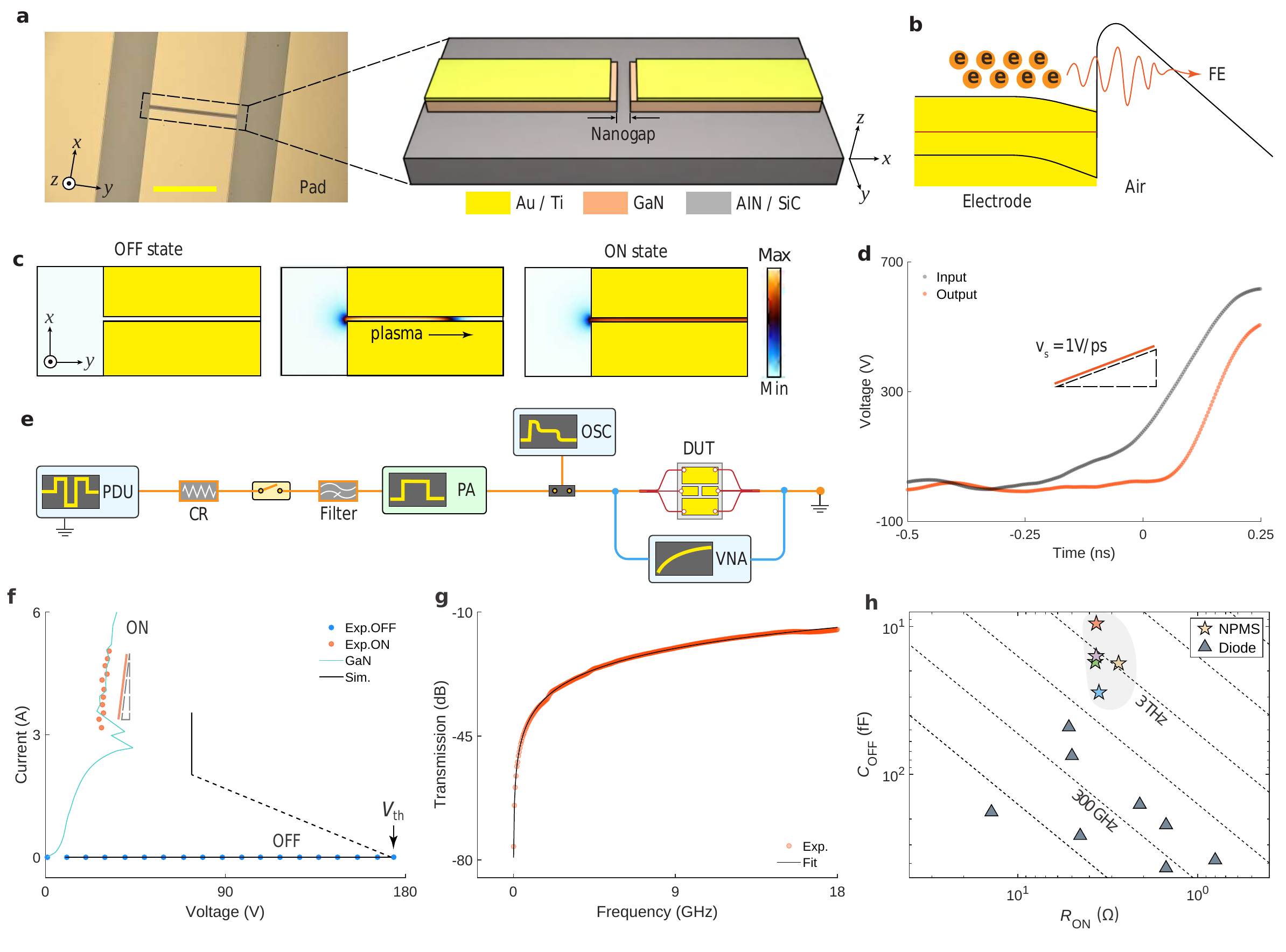"}
	\caption{ \textbf{characterizations of single NPMS device at OFF and ON states}. \textbf{a,} Optical (left) and schematic (right) images of the proposed NPMS fabricated with a coplanar waveguide. Scale bar: \SI{20}{\micro\meter}. \textbf{b,} Energy band diagrams in the electrode-to-air channel direction, where field emission (FE) dominates the plasma formation. \textbf{c,} Simulated distribution of electron density in NPMS during its field-driven transition. \textbf{d,} Measured output waveform showing a 2.1~\si{V ps^{-1}} speed of the device under test (DUT) as high-power signal inputs. \textbf{e,} Schematic of RF probe setup. PDU, power distribution unit; CR, charging resistor; OSC, oscilloscope; PA, power amplifier; VNA, vector network analyzer. \textbf{f,} Measured $I$-$V$ characteristics (points) of the DUT at OFF and ON states, which are matched with simulation and a direct measurement on GaN electrodes, respectively. \textbf{g,} Measured transmission coefficient versus frequency at OFF state, allowing to extract the cut-off capacitance $C_{\mathrm{OFF}}$ of the DUT by fitting. \textbf{h,} Obtained $C_{\mathrm{OFF}}$ versus $R_{\mathrm{ON}}$ for the proposed NPMS devices compared with commercial diodes.}
	\label{fig:2}
\end{figure}

\textbf{Electrical characteristics of on-chip nanoplasma switches.} We first design single NPMS exhibiting tunable impedance as a function of excitation magnitude. As shown in Fig.~\ref{fig:2}a, a 900~\si{nm}-thick n‑doped GaN layer is grown on a semi‑insulating SiC substrate and etched to form a parallel-plate nanogap in the center (Supplementary Note 4). Atomic force microscopy in tapping mode confirmed the well‑defined morphology of the nanogap. The device under test (DUT) is designed as a coplanar waveguide with Ti/Au pads for the electrical characterization on RF-probe platform. We investigate five NPMS devices in total, including asymmetric shapes with triangle electrodes (Supplementary Note 4). The NPMS is normally isolated by air, while as a threshold voltage $V_{\mathrm{th}}$ applied, field emission (FE) triggers gas breakdown within the nanogap\cite{LovelessGarner-108,Yimeng}, generating a highly conductive plasma that switches the device to the ON state (Fig.~\ref{fig:2}b). To elucidate this transient process, we developed a time‑domain simulation model incorporating Fowler–Nordheim theory (Method, Supplementary Note 5). The model relates the internal geometry and material properties of the device to its electrical response, providing, to our knowledge, the first design‑oriented quantitative framework for NPMS. We further discuss the tuning of switching speed and threshold voltage with respect to EF factor and work function (Supplementary Note 5). Simulation results show that high‑density plasma is initially generated near the right‑angle edge and expands across the entire nanogap within a few picoseconds, driven by point‑discharge effects under the enhanced local field (Fig.~\ref{fig:2}c). We then measure the input and output pulses as a function of time (Fig.~\ref{fig:2}d). The slope of the measured curve yields a switching speed $v_{\mathrm{s}}$ of 2.1~\si{V ps^{-1}} for the DUT, which is comparable to the state‑of‑the‑art results from literature\cite{SamizadehNikooJafari-48,ZhaoHuang-47,ZhaoTang-8}. 

We then focus on the performance of NPMS devices in terms of OFF‑state capacitance $C_{\mathrm{OFF}}$ and ON‑state dynamic resistance $R_{\mathrm{ON}}$. To obtain the current‑voltage ($I$-$V$) characteristics, we built a transmission‑line pulse (TLP) test platform (Fig.~\ref{fig:2}e, Method). A pulse generator applies high-voltage pulses to the DUT, while the transient output voltage $V_{\text{out}}$ and $I_{\text{out}}$ waveforms are recorded using a digital oscilloscope (25~\si{GS/s} sampling rate, 1~\si{GHz} bandwidth). The pulse has a width of \(100~\mathrm{ns}\) and a rise time of \(10~\mathrm{ns}\). When the pulse amplitude reaches the threshold, the DUT turns on and passes the response waveform to the 50~$\Omega$ termination of the oscilloscope. As shown in Fig.~\ref{fig:2}f, the measured $I_{\text{out}}$ remains near zero and then increases sharply to 3.2~\si{A}, corresponding to a voltage drop from 174.2 to 28.1~\si{V}. The on-state $I$-$V$ data yields an $R_{\mathrm{ON}}$ of 2.8~\si{\Omega} for the DUT, which deviates considerably from the simulation. Tests on five NPMS devices reveal that the measured $I$-$V$ characteristics are, in fact, consistent with direct short-circuit measurements on GaN electrodes (Supplementary Note 6). We thus attribute the total $R_{\mathrm{ON}}$ of NPMS to the intrinsic ohmic resistance of the GaN electrode. Furthermore, the transmission $S_{21}$ versus frequency is measured using a vector network analyzer directly connected to the DUT, which is fitted to extract a $C_{\mathrm{OFF}}$ of 17.3~\si{fF} (Fig.~\ref{fig:2}g). The resulting cut-off frequency of NPMS devices, defined as $1/2\pi(C_{\mathrm{OFF}}\times R_{\mathrm{ON}})$\cite{Mohammad}, is approximately an order of magnitude higher than that of commercial diodes (Fig.~\ref{fig:2}h and {Extended Table~2}), indicating the potential of NPMS for RF front-end protection in a high-frequency regime.\\

\begin{figure}
	\centering
	\includegraphics[width=1\linewidth,angle=0]{"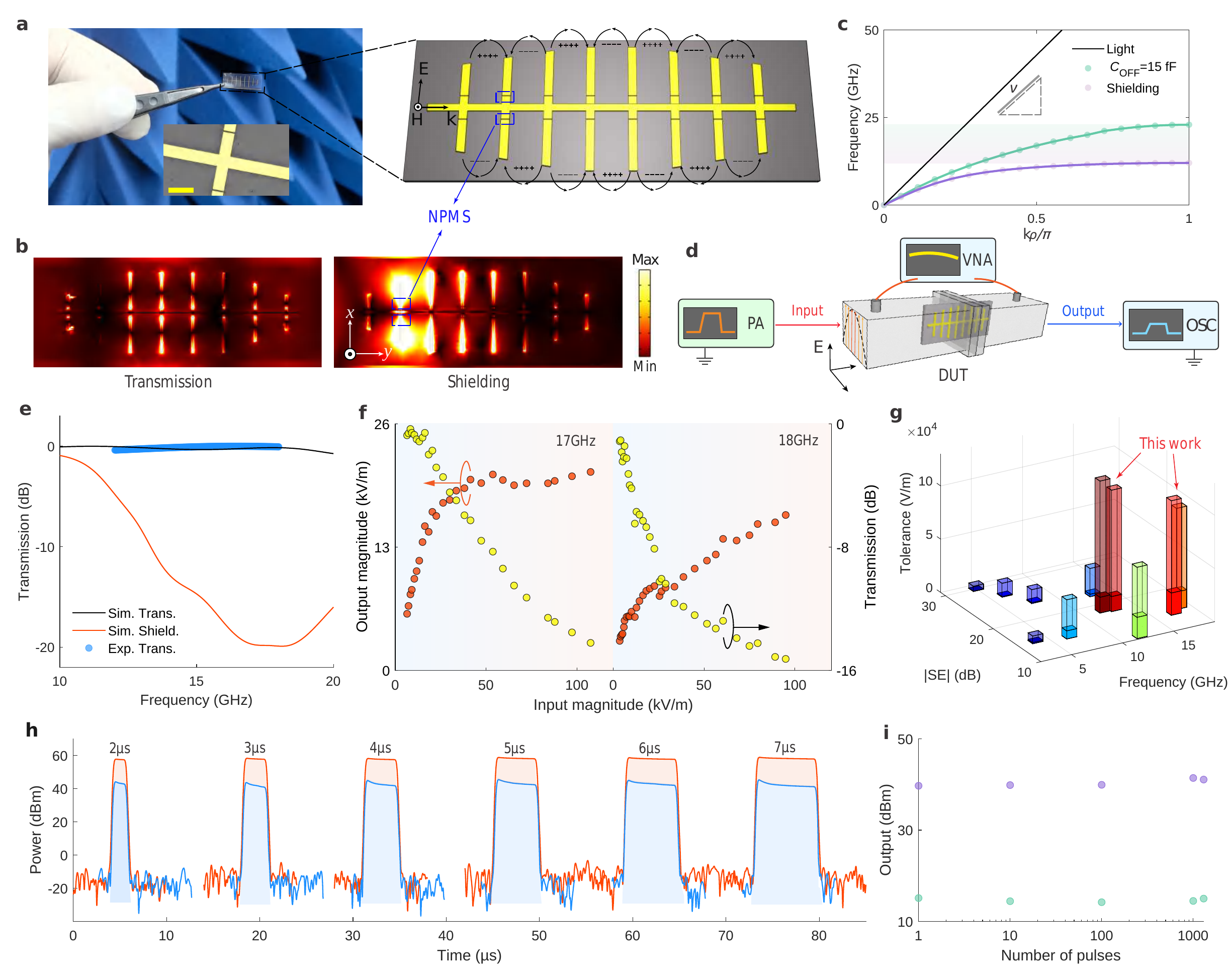"}
	\caption{ \textbf{Spoof surface plasmon polariton metasurface (SPPM) loaded with NPMS.} \textbf{a,} Schematic (left) and parallel incident wave excitation of SPPM. Scale bar: \SI{20}{\micro\meter}. \textbf{b,} Simulated field distribution in transmission and shielding mode, showing strong localized enhancement at NPMS that actuates switching. \textbf{c,} Calculated dispersion relation of the SPPM; the slope of the curve defines the group velocity $\nu$. \textbf{d,} Schematic of waveguide injection setup. \textbf{e,} Transmission coefficient $S_{21}$ as a function of frequency; drawn lines, full-wave simulation; dots, measured results in transmission mode. \textbf{f,} Measured output magnitude (left axis) and corresponding $S_{21}$ (right axis) under high-power excitation at 17 and 18~\si{GHz}, respectively. \textbf{g,} Comparison of operating band, shielding effectiveness (SE) and tolerance field among SPPM and diode-based metasurfaces. The solid color of bars represents the field magnitude after limiting. \textbf{h,} Measured input and output waveforms in the time domain for pulse widths ranging from 2 to 7~\si{\micro\second}. \textbf{i,} Stability test under repetitive pulse excitation (1,300 cycles), showing the variation in output power for transmission and shielding modes.
     }
	\label{fig:3}
\end{figure}

\textbf{Design of nonlinear metasurface based on NPMS.} Metasurfaces enable efficient spatial‑wave modulation with the advantages of low mass and low power consumption. Here we design a spoof surface plasmon polariton metasurface (SPPM) integrated with on‑chip NPMS to achieve the field-driven adaptivity (Fig.~\ref{fig:3}a). The device consists of a symmetrical fish-bond transmission line, where the NPMS are loaded on each branch. The $y$-polarized field drives free electrons to gather toward the branch edges and excite plasmon. When NPMS is OFF, impedance matching between the device and free space permits low‑loss transmission of normal signals; while as NPMS triggered by a high-power field, the increased electrical length of each branch disrupts the matching condition, leading to a strong suppression on incident waves. Electric field distribution in SPPM obtained by full-wave simulation gives a direct insight into this adaptive performance (Fig.~\ref{fig:3}b). The underlying mechanism is further illustrated by the calculated dispersion relation of SPPM (Fig.~\ref{fig:3}c and Supplementary Note 7). The group velocity $\nu$, given by the slope of dispersion, gradually decreases with frequency and approaches zero at the cut‑off frequency $f_{\mathrm{c}}$ \cite{JiangDeng-3,ZuWu-225}. Therefore, the operational bandwidth of SPPM is bounded by $f_{\mathrm{c}}$ between transmission and shielding modes. 

We evaluate the field‑driven adaptive response of the proposed SPPM by waveguide injection tests (Fig.~\ref{fig:3}d and Methods). Structural parameters and the conversion from injected power to internal electric field in the waveguide are detailed in Supplementary Note 8. In transmission mode, the measured $S_{21}$ exhibits an insertion loss (IL) below 0.4~\si{dB} across 12 to 18~\si{GHz}, in good agreement with full‑wave simulations (Fig.~\ref{fig:3}e). High‑power injection is then tested at discrete frequencies. At 17~\si{GHz} and 18~\si{GHz}, the output fields are clamped to $2.1\times10^4$ and $1.6\times10^4$~\si{V/m}, respectively, corresponding to shielding effectiveness (SE) values of approximately 14.2 and 15.3~\si{dB} (Fig.~\ref{fig:3}f). The progressive increase in SE with input power cannot be captured by the transient impedance variation of the single NPMS. Rather, as the input field increases, the integrated NPMS elements along the propagation direction ($x$-axis) turn on sequentially (Fig.~\ref{fig:3}b), producing the continuous SE characteristic observed. A second SPPM designed for the 10–13~\si{GHz} band exhibits similar EM characteristics (Supplementary Note 8). Fig.~\ref{fig:3}g compares the performance of SPPM with that of diode-based metasurfaces, demonstrating a roughly twofold enhancement in tolerance field (from 67 to 123~\si{kV/m}) while effectively suppressing HPMs to intensities within the operable range of conventional protection (Extended Table~3). Owing to the on‑chip integration that eliminates parasitic effects, the operating band can be extended to the millimeter‑wave frequencies without sacrificing SE. 

We also fabricate a 5$\times$4 array of SPPM elements and characterize its protection performance under spatial irradiation in a microwave anechoic chamber (Supplementary Note 8). As expected, the obtained transmission curves agree well with full-wave simulation. In addition, benefiting from the picosecond switching speed of NPMS, the SPPM delivers smooth output waveforms without observable spike leakage and can withstands high‑power signals across pulse widths from 2 to 7~\si{\micro\meter} (Fig.~\ref{fig:3}h). To assess long‑term reliability against potential degradation from electro-migration and ion bombardment, we perform 1300 repetitive pulse tests (pulse width: 200~\si{ns}; magnitude: $5.1\times10^4$~\si{V/m}) on the SPPM. The output powers in transmission and shielding mode vary by only 3.3~$\%$ and 0.8~$\%$, respectively, confirming stable adaptivity under repeated pulsed operation.\\

\begin{figure}
	\centering
	\includegraphics[width=1\linewidth,angle=0]{"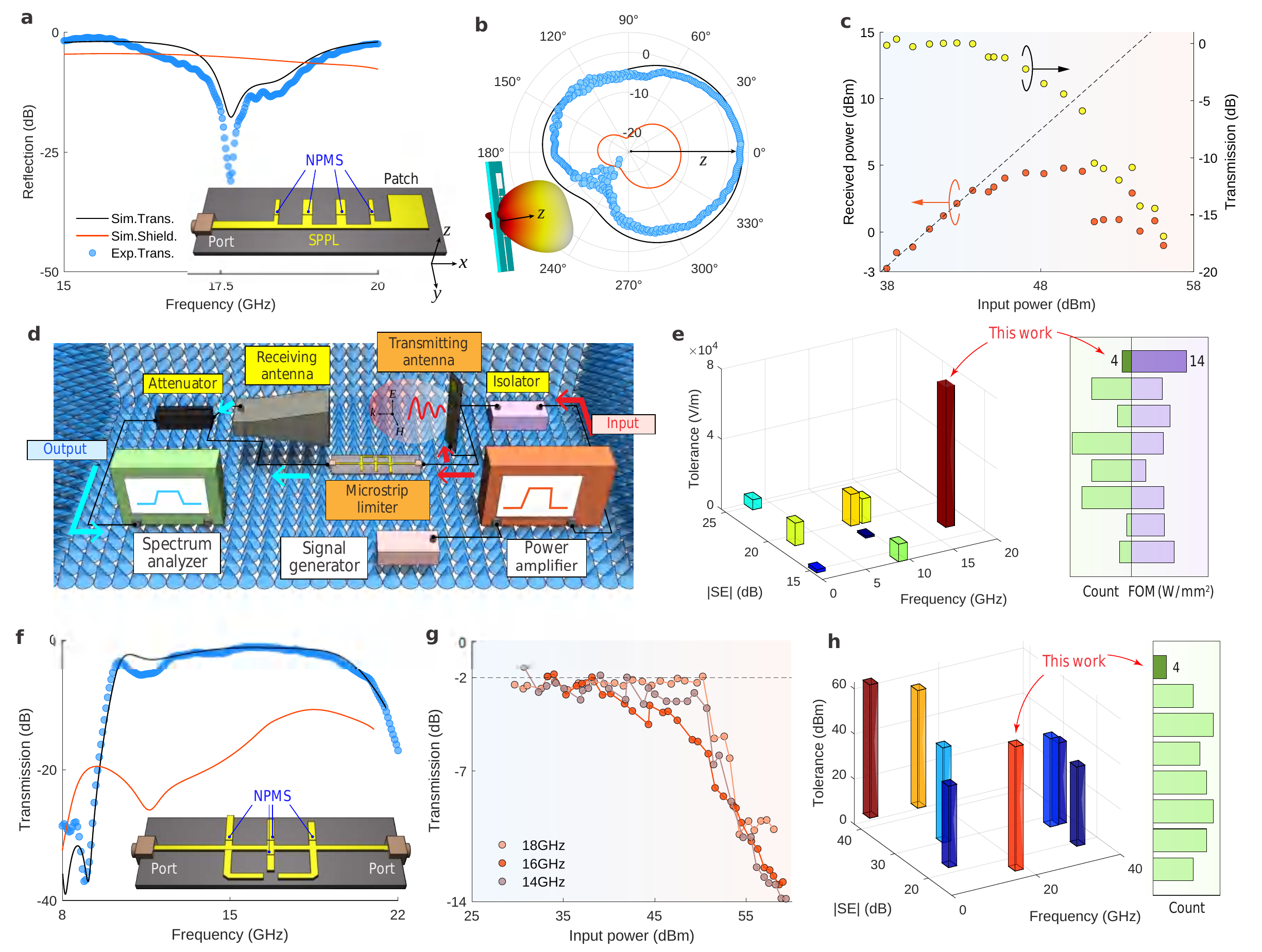"}
	\caption{ \textbf{Microstrip patch antenna and circuit limiter for adaptive EM protection.} \textbf{a,} Reflection coefficient $S_{11}$ versus frequency for the proposed antenna in transmission and shielding modes; insert, schematic model. \textbf{b,} Radiation patterns at the central frequency of 17.5~\si{GHz}; insert, full-wave simulation. Drawn lines and dots follow the same convention as in \textbf{a}. \textbf{c,} Measured received power (left) and transmission coefficient $S_{21}$ (right) as the function of input power; dashed line, linear fitting to transmission mode. \textbf{d,} Measurement setup inside the microwave anechoic chamber for high‑power injection testing of the antenna and limiter. \textbf{e,} Left, comparison of operating band, shielding coefficient (SE) and tolerance between the proposed antenna and diode-based devices; right, switches count and the figure of merit (FOM) for tolerance power density. \textbf{f,} $S_{21}$ versus frequency for the proposed limiter; insert, schematic model. \textbf{g,} Measured $S_{21}$ as the function of input power at different frequencies. \textbf{h,} Left, Same comparison as in \textbf{e} (left) and switches count (right) for the proposed limiter.
    }
	\label{fig:4}
\end{figure}

\textbf{Protective antenna and circuit limiter devices.} 
We further design a protective microstrip patch antenna which adopts a spoof surface plasmon polariton transmission line (SPPL) configuration with NPMS loaded on its stubs to modulate the frequency response (Fig.~\ref{fig:4}a, insert). When the NPMS is OFF, the radiation band of the antenna falls within the passband of the feeding line, allowing received signals to propagate towards the port; upon excitation by high-power signals, the NPMS turn ON, converting the original passband into a stopband and thus blocking the transmission between the port and the radiating patch to achieve protection. The simulated port reflection coefficient $S_{11}$ of the antenna is shown in Fig.~\ref{fig:4}a. In transmission mode, $S_{11}$ remains below -10~\si{dB} around the design frequency 17.5~\si{GHz}, demonstrating a good agreement with the results measured by VNA. While in shielding mode, $S_{11}$ rises to roughly -5~\si{dB}, indicating suppressed signal radiation and reception. The radiation patterns at 17.5~\si{GHz} show a drop in main‑lobe gain from 3.4~\si{dBi} to ‑11.9~\si{dBi} (Fig.~\ref{fig:4}b), a reduction of 14.5~\si{dB} that validates the efficient attenuation of external high‑power coupling, with the measured patterns and 3~\si{dBi} gain in transmission mode following the same trend.

To evaluate the protection performance under high‑power signals, we adopt the proposed antenna as a transmitter and a standard horn antenna as the receiver inside a microwave anechoic chamber (Fig.~\ref{fig:4}d and Method). In low‑power regime, the received power increases linearly with input power, indicating normal operation with the NPMS remaining OFF (Fig.~\ref{fig:4}c). Beyond a triggering threshold of 44~\si{dBm}, the received power saturates at approximately 5~\si{dBm}, denoting the onset of adaptive protection as NPMS turns ON, and then gradually decreases to -1~\si{dBm}, corresponding to a SE of about 16~\si{dB}. From the antenna reciprocity principle, we extract a tolerance field strength of 8$\times 10^4$~\si{V/m} for the proposed antenna (Supplementary Note 8). Compared with reported designs, our approach offers a higher operating frequency, comparable SE and substantially improved tolerance to HPMs (Fig.~\ref{fig:4}e and Extended Table~4). Notably, with only four NPMS elements, the normalized power density figure-of-merit (FOM)—evaluated in terms of geometrical size, effective aperture and gain—is also considerably enhanced relative to other microstrip and reflector-type antennas (Supplementary Note 8).

Figure~\ref{fig:4}f depicts the microstrip limiter incorporating NPMS. In transmission mode, the device exhibits bandpass filtering with IL below 2~\si{dB} from 13.3 to 18.6~\si{GHz}, where the measured $S_{21}$ between the ports matches well with simulation. Upon the trigger of NPMS, we observe an SE exceeding 10~\si{dB}. The protection response under high-power injection is then characterized at different frequencies (Fig.~\ref{fig:4}g). The transmission $S_{21}$ remains flat at low input levels and rolls off once the power surpasses the triggering threshold. At an input of approximately 58~\si{dBm}, the SE exceeds 10~\si{dB} across the entire measured band, indicating effective limiting action. By virtue of the high cut-off frequency and power capacity of NPMS, the proposed limiter achieves robust protection over the Ku band with a compact on‑chip integration of switch elements and EM topology (Fig.~\ref{fig:4}h and Extended Table~5). Relative to conventional semiconductor limiters, this design offers extended frequency agility, simplified fabrication and enhanced power handling within a decreasing footprint.\\

Reducing the switching threshold voltage of NPMS could markedly extend their lifetime under high‑power irradiation by alleviating thermal and electrical stress on the electrodes. We show that introducing multiple nanotips on the electrode surface or reducing the nanogap scale can substantially enhance the field‑enhancement factor $\beta$, which describes the electron emission efficiency of the cathode, leading to an order‑of‑magnitude reduction in threshold voltage (Supplementary Note 5). Such modifications, however, demand advanced lithography and etching techniques with high aspect ratios, particularly for large‑area fabrication. Alternatively, selecting electrode materials with low work function $\phi$, excellent thermal stability or strong resistance to ion sputtering offers another route to improved long‑term reliability. Graphene, for instance, combines the highest known thermal conductivity with a sublimation temperature of 3700~\si{K} under an inert atmosphere, making it suitable for high‑tolerance electrodes with ultrafast heat dissipation\cite{Alexander,Balandin}. Lanthanum hexaboride (LaB$_6$) also emerges as a promising candidate, owing to its low $\phi$ (below 2.5~\si{eV}) that reduces turn‑on threshold and high melting point (approximately 3000~\si{K}) that enhances device longevity\cite{ZhangTang-276,ZhangJimbo-278}. Its susceptibility to oxidation in air, however, raises $\phi$ and causes structural degradation, highlighting the importance of an inert gas environment for practical implementation. Beyond material selection, the adhesion strength between electrodes and substrates requires further attention to prevent thermal-stress-induced cracking, peeling and other failure modes of thin-film electrodes.

Beyond the Ku‑band designs demonstrated in this work, on‑chip nanoplasma also holds promise for EM protection at higher frequencies. The measured NPMS exhibits a rated operational current versus cut‑off frequency product ($I_{\text{ro}} \times f_{c}$) that is nearly two orders of magnitude higher than that of commercial diodes (Supplementary Note 6). This addresses a fundamental limitation of PIN diodes, which suffer from an inherent trade-off between switching speed and power tolerance. In contrast, NPMS is triggered directly by field emission and is not constrained by carrier mobility, circumventing this trade‑off altogether. Another promising direction is integrating nanoplasma structures onto flexible substrates, as the deposition of conductive materials like metals and 2D materials on such stages is practically feasible\cite{Rogers, ShanJiang}. We verify that bending or stretching of the flexible substrate induces only negligible deformation of the nanogap, with no discernible effect on the electrical characteristics of the device (Supplementary Note 9). These attributes, together with stability and inherent scalability of on‑chip integration, collectively make nanoplasma well suited for EM protection across diverse scenarios and various types of information platforms.\\

This work presents that nanoplasma switches overcome the intrinsic limitations of conventional semiconductor switches, substantially extending the operating band and tolerance power of adaptive EM protectors in RF front end. The high performance of our proposed devices demonstrated here hold promise for future studies on EM compatibility and protection, enabling reliable responses across multiple coupling pathways while extending coverage to millimeter‑wave and terahertz bands. More broadly, the on‑chip nanoplasma concept offers a versatile route for efficient tunability across a wide range of functional devices, including programmable metasurfaces, energy harvesters, opto-electronic chips and pulse-width modulators, with material breadth spanning wide‑bandgap semiconductors, inorganic metallic compounds and 2D materials. These advances open new avenues for next‑generation RF front ends and emerging THz systems.

\section*{Methods}
 
\textbf{Fabrication of on-chip nanoplasma switches.} NPMS were fabricated using a two-step photolithography process. First, the GaN wafer was treated with hexamethyldisilazane (HMDS) coated with AR80 photoresist and then patterned using a stepper with an exposure time of \(300~\mathrm{ms}\). After development in a \(2.38\%\) tetramethylammonium hydroxide (TMAH) solution and optical inspection, the photoresist was hard-baked and subjected to oxygen-plasma descumming at \(200~\mathrm{W}\) for \(2~\mathrm{min}\). The exposed GaN was subsequently removed by ion-beam etching at an energy of \(400~\mathrm{eV}\) and a beam current of \(100~\mathrm{mA}\) for \(90~\mathrm{min}\). The remaining photoresist was stripped by ultrasonic cleaning in acetone and isopropyl alcohol, followed by rinsing with deionized water and nitrogen drying. A second photolithography process was then performed to define the metal electrodes. After HMDS treatment, 1303 photoresist was spin-coated, exposed for \(4~\mathrm{s}\) using a mask aligner, post-exposure baked at \(110\,^{\circ}\mathrm{C}\) for \(2~\mathrm{min}\) and developed in a \(2.38\%\) TMAH solution for \(40~\mathrm{s}\). A Ti/Au metal stack with thicknesses of \(20~\mathrm{nm}\) and \(100~\mathrm{nm}\), respectively, was deposited by magnetron sputtering. The electrode patterns were formed by lift-off in acetone and isopropyl alcohol, after which the sample was rinsed, nitrogen-dried, and diced along the predefined alignment marks. All the proposed NPMS-based designs were also fabricated by the same process.

\textbf{TLP tests to determine resistance and capacity.} Pulsed measurement minimizes Joule heating and captures the transient electrical response of the device. For each pulse, the voltage and current values are extracted from the flat top of the recorded waveforms and averaged to generate a single data point on the $I-V$ curve. The pulse amplitude is then increased in steps and the measurement repeated to obtain the full $I-V$ characteristic. From the slope of the ON‑state results, the dynamic ON‑resistance $R_{\text{ON}}$ is extracted. The results also show that NPMS sustains ON-state currents of several amperes and dissipates over one hundred watts without failure, indicating a power capacity that substantially exceeds that of conventional semiconductor diodes.

\textbf{Numerical simulations.} To elucidate the transient response process of the NPMS, we build the time-domain simulation model incorporating Fowler-Nordheim theory in COMSOL Multiphysics software by coupling the electric discharge module and circuit module. The Fowler–Nordheim theory was employed to characterize the FE process between electrodes. The simulations of all proposed NPMS-based designs, including $S$-parameters and field distributions, were carried out by CST full-wave software. The electrical characteristics of NPMS at ON and OFF states were equivalent to a resistor and a capacitor, respectively.

\textbf{Waveguide injection tests.} We evaluate the field‑driven adaptive response of the proposed SPPM by waveguide injection tests. The SPPM prototype is placed in the middle of the WR62 waveguide with cross-section dimensions of 15.799~\si{mm} $\times$ 7.899~\si{mm} (see Fig.~\ref{fig:3}g). For low-power measurement, the input and output ports of the waveguide are directly connected to a vector network analyzer (VNA) after calibrating. The output power of the VNA is set to -20~\si{dBm} to provide a low EM field environment for the SPPM prototype and measure its transmission coefficient versus frequency. For high-power injection, the microwave signal is generated by a signal generator (SG) and then amplified by a power amplifier (PA). Subsequently, the amplified waves are injected into the input port of the waveguide, which can provide enough EM field in the waveguide to induce the SPPM prototype (Supplementary Note 8). At the output port, a spectrum analyzer (SA) is connected to monitor the transmitted power through SPPM prototype. System losses are calibrated prior to measurements to ensure accurate power quantification. The shielding effectiveness (SE) is calculated by SE~(dB) = $P_\mathrm{in}\mathrm{(dBm)} - P_\mathrm{out}\mathrm{(dBm)}$, where $P_\mathrm{in}$ and $P_\mathrm{out}$ represent the input and output power of the waveguide, respectively.

\textbf{Antenna protection performance tests.} We adopt the proposed antenna as a transmitter and a standard horn antenna as the receiver inside a microwave anechoic chamber (Supplementary Note 8). By increasing the input power, we monitor the received power and characterize the transmission of the antenna. The decrease of received power as input increases indicates the onset of adaptive protection, where SE can be quantified according to the deviation between transmission and reception. Owing to antenna reciprocity, the change in receiving capability is directly inferred from the measured radiation response, enabling a convenient evaluation of protection. Compared with conventional HPM irradiation testing\cite{Zhasong}, our approach imposes less demanding requirements on the frequency band and power level of the high‑power source, and offers greater experimental flexibility\cite{FangWu-308}.

\section*{Extended Data}

\begin{table}
\centering
\caption{Operating frequencies and generated field strength of state-of-the-art pulsed HPM sources.}
\label{tab:Extended Table 1}
\renewcommand{\arraystretch}{2}
\scriptsize
\setlength{\tabcolsep}{5pt} 
\begin{tabular}{@{}lcccccccc@{}}
\hline
\textbf{Freq. (GHz)} &1.31  & 1.18  &1.10  &  1.20 &  2.15& 2.80  & 2.61   & 3.53  \\
\textbf{Power (GW)}  & 15.07  &  7.19&3.63   & 1.78  &   3.09&  2.99 &0.95  &1.68 \\
\textbf{Peak Field (kV/m)}&300.65 & 207.64&147.66 &103.45 & 136.07&134.04 &75.38 & 100.29\\
\hline
\textbf{Freq. (GHz)}  &3.63 & 3.90  &3.90  &9.45  &  9.45& 31.06  & 46.01   & 60.16\\
\textbf{Power (GW)}  &5.00  &  4.48 &3.75  & 2.94 & 14.75& 2.97  &   3.46 &0.99   \\
\textbf{Peak Field (kV/m)}&173.18 & 163.90& 150.09&132.80 & 297.52&133.52 & 144.12& 76.96\\
\hline
\textbf{Freq. (GHz)} & 140.23 &  16.97 & 9.56 &8.73  & 8.25 &17.03   & 35.05   & \\
\textbf{Power (GW)} & 1.86 &0.99   & 1.10 & 0.41 & 0.30 &  0.47 & 0.95   &  \\
\textbf{Peak Field (kV/m)}& 105.73&77.11 & 81.25& 49.35& 42.31& 52.86&75.35 & \\
\hline
\end{tabular}
Note: Field strength is calculated via Friis free-space transmission formula, assuming a transmitting gain of 30~\si{dBi} at an operating distance of 100~\si{m}. The database of pulsed HPM sources is obtained from literature\cite{2024HighPowerMicrowaves}. 
\end{table}

\begin{table}
\centering
\caption{Electrical characteristics of diodes and NPMS.}
\label{tab:Extended Table 2}
\renewcommand{\arraystretch}{2}
\scriptsize
\setlength{\tabcolsep}{5pt} 
\begin{tabular}{@{}lccccccccc@{}}
\hline
\textbf{Diode No. } & 1 &2  & 3 &   4 &  5&  6   & 7 &   8 & This work    \\ 
\hline
\textbf{{$R_{\mathrm{ON}}$} $(\Omega)$}&1.5  &14  & 5.2 & 1.5  &  0.8&  2.1& 4.5 & 5& 2.8 \\
\textbf{$C_{\mathrm{OFF}}$(fF)}  & 430 & 180 &48  &  220&  380&160  & 260 &75 & 17.3   \\
\textbf{Cut-off Freq. (GHz)}& 248.8 &  63.2&  637.6& 482.3&523.5&473.7 &136.1 & 424.4& 3349.2 \\
\hline
\end{tabular}
Note: Diodes No.1 to 8 correspond to the commercial diodes of BAP51-02, NSR201MX-D, MA4AGFCP910, SMP1345, MMP4401, MA4L011-134, NSVP249SDSF3, MA4SPS402, respectively, where a general parasitic capacitance of 30~\si{fF} is considered. The cut-off frequency is calculated by ${1}/{2 \pi R_{\mathrm{ON}} C_{\mathrm{OFF}}}$\cite{Mohammad}.
\end{table}

\begin{table}
\centering
\caption{Comparison of diode-based nonlinear metastructures and the proposed SPPMs.}
\label{tab:Extended Table 3}
\renewcommand{\arraystretch}{2}
\scriptsize
\setlength{\tabcolsep}{5pt} 
\begin{tabular}{@{}lcccccccc@{}}
\hline
\textbf{Ref. } &\cite{WuLiu-368} &\cite{JiangDeng-3}  &\cite{WuXu-148} &  \cite{GongZhang-370} &  \cite{TianHuang-369}  &  \cite{JiangDeng-366} & This work     \\
\hline
\textbf{Freq. (GHz)}  &12 &6.8  & 13& 3.8  & 8 / 6.5    &5 & 12 / 13 / 17 / 18\\
\textbf{SE (dB)}  &10.7 &14.3  &22.7 &15    &   24.2 / 27.2    &30 & 18.5 / 18.5 / 14.2 / 15.3\\
\textbf{Tolerance field (kV/m)}  & 67&38  &26.5 &  7.2 &13 / 13  &4.9 & 123.9 / 114 / 107.5 / 94.9  \\
\textbf{Leakage field (kV/m)}  &20 &7.6  &2.33 & 1.05  & 0.8 / 0.56   &0.15& 14.8 / 14.1 / 20.9 / 16.4\\
\hline
\end{tabular}
\end{table}

\begin{table}
\centering
\caption{Comparison of diode-based protective antennas and the proposed protective antenna.}
\label{tab:Extended Table 3}
\renewcommand{\arraystretch}{2}
\scriptsize
\setlength{\tabcolsep}{3pt} 
\begin{tabular}{@{}lcccccccc@{}}
\hline
\textbf{Ref. } &\cite{DengLin-375} &\cite{FangWu-376}  &\cite{Zhasong} &  \cite{QuZha-378} & \cite{QuZha-379} &\cite{WangTang-380}   &  \cite{ZhaLiu-374} & This work     \\
\hline
\textbf{Freq. (GHz)}  &2.3 &0.8  & 9.5&10   & 10  & 3.1   &10.8 &17.5 \\
\textbf{SE (dB)}  &12.9 &1.8 &18 &   1 &  10& 5.7  &14 &  16 \\
\textbf{Tolerance field (kV/m)}  & 19.3&15 &20 &  18 &14   &  25  &20 & 80.2 \\
\textbf{Switch number}  &6 & 2 &1660 & 388 &7200  & 8 &380& 4 \\
\textbf{Gain (dBi)}&2 & 2 &25  &27.4 & 20.9  & 2.14 &28   &3 \\
\textbf{FOM (W/mm$^2$)}& 3.32$\times 10^{-1}$  & 1.81$\times 10^{-2}$ & 1.71$\times 10^{-2}$  & 7.12$\times 10^{-5}$ & 1.37$\times 10^{-2}$ & 1.05$\times 10^{-1}$ &  1.01$\times 10^{-1}$ & 14.0 \\
\hline
\end{tabular}
Note: The normalized power density figure-of-merit (FOM) is calculated by the operating frequency, geometrical size, effective aperture and gain of the antennas (Supplementary Note 8).
\end{table}

\begin{table}
\centering
\caption{Comparison of diode-based limiters and the proposed on-chip limiter. }
\label{tab:Extended Table 4}
\renewcommand{\arraystretch}{2}
\scriptsize
\setlength{\tabcolsep}{5pt} 
\begin{tabular}{@{}lcccccccc@{}}
\hline
\textbf{Ref. } &\cite{LiMa-358} &\cite{ZhangZhou-359}  &\cite{HaoGu-360} & \cite{YangZhang-361}  & \cite{ZhuWang-362} &\cite{WuKang-363}   &\cite{LiWang-364}   & This work     \\
\hline
\textbf{Freq. (GHz)}  &18 &2  & 36&36   & 35  & 10   &12 &18 \\
\textbf{SE (dB)}  &40.5 &41 &24.9 &   17 &  25  &20 &  28 &14\\
\textbf{Tolerance power (dBm)}  & 56&63 &43 &  39 &40  &  40  &46 & 59 \\
\textbf{Switch number}  &12& 16 &18 & 16 &14  & 18 &12& 4 \\
\hline
\end{tabular}
\end{table}

\backmatter

\bmhead{Supplementary information}

Any methods, additional references, Nature Portfolio reporting summaries, source data, extended data, supplementary information, acknowledgements, peer review information; details of author contributions and competing interests; and statements of data and code availability are available at XXXXX.

\bmhead{Acknowledgements}

The authors wish to acknowledge Qian Dong, Yuan Xu, Yunquan Mei and Xiaocheng Ni for their helpful suggestions on the manuscript and their support of our project. The authors also want to thank Mingtuan Lin, Huan Jiang, Zhuang Qu,Tao Tian, Kui Wen and Mo Li for the helpful discussions. Part of the high-power radiation experiments were done at Hunan Vanguard Group Co., Ltd. This work had received funding from National Natural Science Foundation of China under Grant No. 62401592 and the Young Scientists Innovation Fund Project of National University of Defense Technology under Grant No.~ZK24-10.

\bibliography{sn-bibliography}

\end{document}


\begin{center}
    \vspace*{1em}
    {\LARGE \textbf{Supplementary Information}} \\[1.5em]
    {\Large \textbf{On-chip nanoplasma for adaptive electromagnetic protection}} \\[1em]
   \end{center}
\vspace{2em}

\tableofcontents
\vspace{3em}


\section{Experiment on an embodied intelligent robot irradiated by HPM}
\begin{figure}[H]
  \centering
  \includegraphics[width=\linewidth]{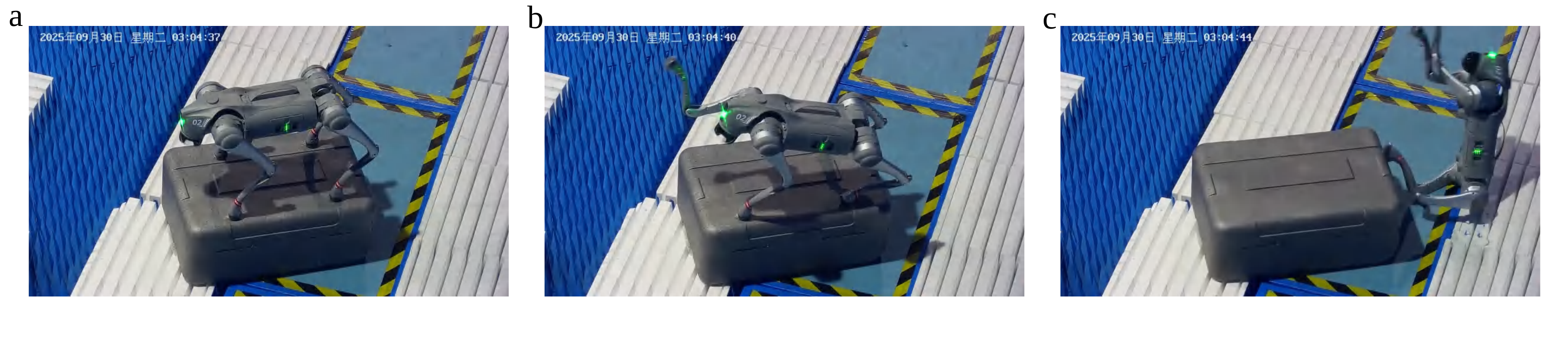}
  \caption{Embodied intelligent robot in \textbf{a,} the initial power-on state, \textbf{b,} the normal operational state, and \textbf{c,} the inoperable state following HPM radiation.}
  \label{fig:robot}
\end{figure}

Initially, the embodied intelligent robot successfully powers on and runs properly (Fig. \ref{fig:robot}a), while performing interactive greeting gestures (Fig. \ref{fig:robot}b). Nevertheless, upon exposure to HPM radiation, 
the robot collapses instantly as if being struck, falling to the ground and suffering permanent functional paralysis (Fig. \ref{fig:robot}c).

\section{Limitations of semiconductor diodes}
\begin{figure}[H]
  \centering
  \includegraphics[width=0.6\linewidth]{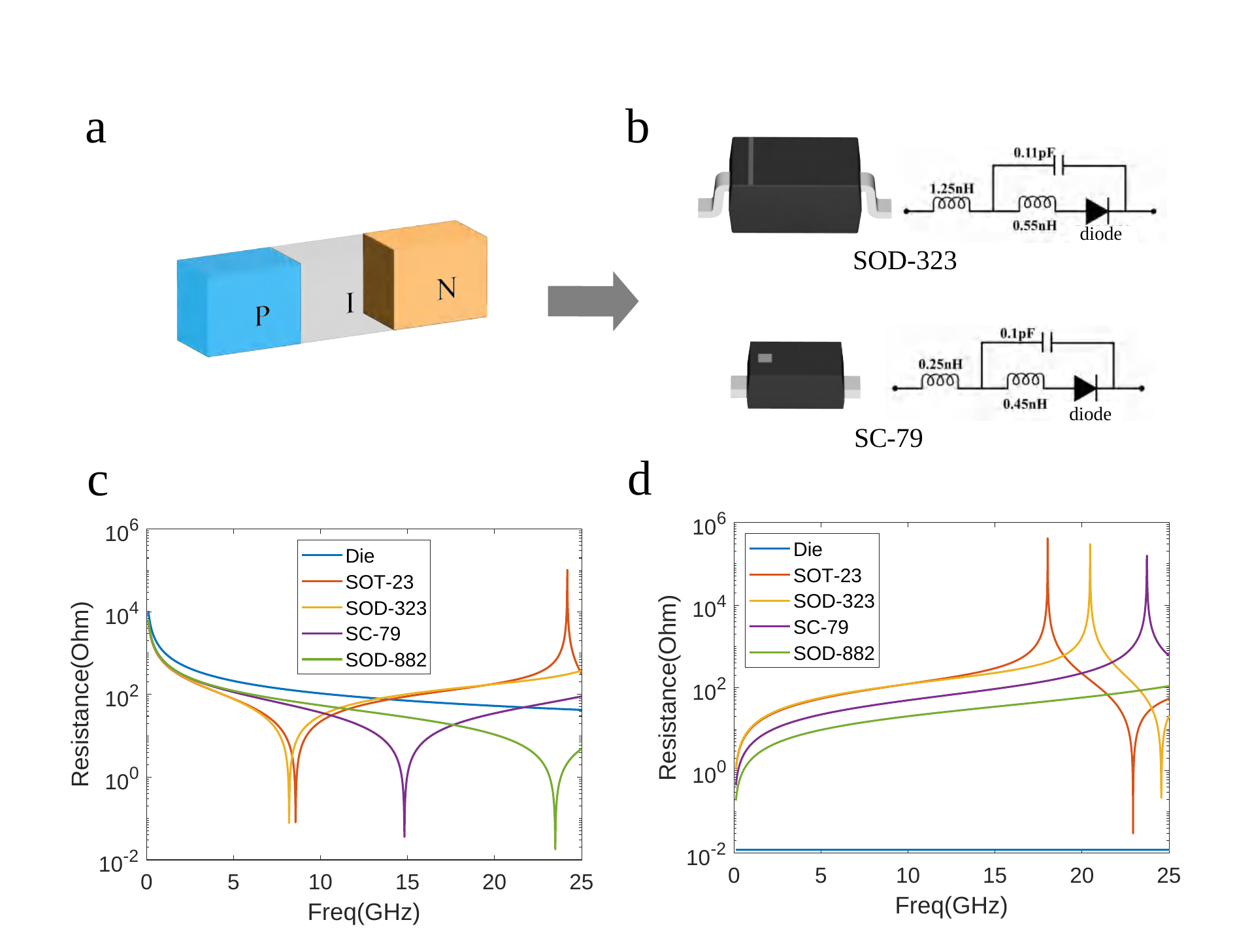}
  \caption{\textbf{a,} Geometric diagram of a PIN diode. \textbf{b,} Packaging and interconnect parasitic effects. \textbf{c,} OFF-state and \textbf{d,} ON-state impedance curves of SMP1320 diodes for various packages.}
  \label{fig:diode}
\end{figure}

The core trade‑off for diodes used in high‑frequency, high‑power radio-frequency (RF) front‑end protection lies in the difficulty of simultaneously achieving low cut-off capacitance, high power‑handling capability, and fast response. Taking the widely used PIN diode in high-power electromagnetic protection as an example: the PIN diode consists of a heavily doped P$^+$ region, an intrinsic or lightly doped I region, and a heavily doped N$^+$ region (Fig. \ref{fig:diode}a). Under zero DC bias, when a high-power microwave signal is injected, carriers are injected from the P$^+$ and N$^+$ boundaries into the I region during the positive half-cycle of the signal. During the negative half-cycle, not all the injected carriers can return; some remain in the I region. After several cycles, a stable non-equilibrium carrier accumulation builds up in the I region, triggering conductivity modulation, which converts the I layer from a high-resistance state to a low-resistance state, thus achieving self-turn-on. For applications at high frequencies, diodes must exhibit low cut-off capacitance. Reducing the junction area decreases cut-off capacitance and improves high-frequency isolation, but compromises current capacity and power handling capability; increasing the junction area enhances power capacity but increases cut-off capacitance. Thickening the I layer reduces capacitance but raises on-resistance and response time. 

At even higher frequencies, performance limitations shift progressively from intrinsic chip parameters to packaging and interconnect parasitic effects—parasitic capacitance and inductance can cause direct conduction in the cut-off state (Fig. \ref{fig:diode}b). Taking the SMP1320 PIN diode manufactured by Skyworks Solutions, Inc. as an example, the impedance curve of the diode in the off state shown in Fig.~\ref{fig:diode}c exhibits high impedance at low frequencies. As frequency increases, the impedance begins to decrease to varying degrees until reaching the resonant frequency, after which it starts increasing again. At the resonant frequency, the cut-off impedance reaches its minimum due to the parasitic effects, and  the impedance of the diode is comparable in the OFF and ON states (Fig.~\ref{fig:diode}d). which implies that the diode loses its switching capability. 

\begin{figure}[H]
  \centering
  \includegraphics[width=0.9\linewidth]{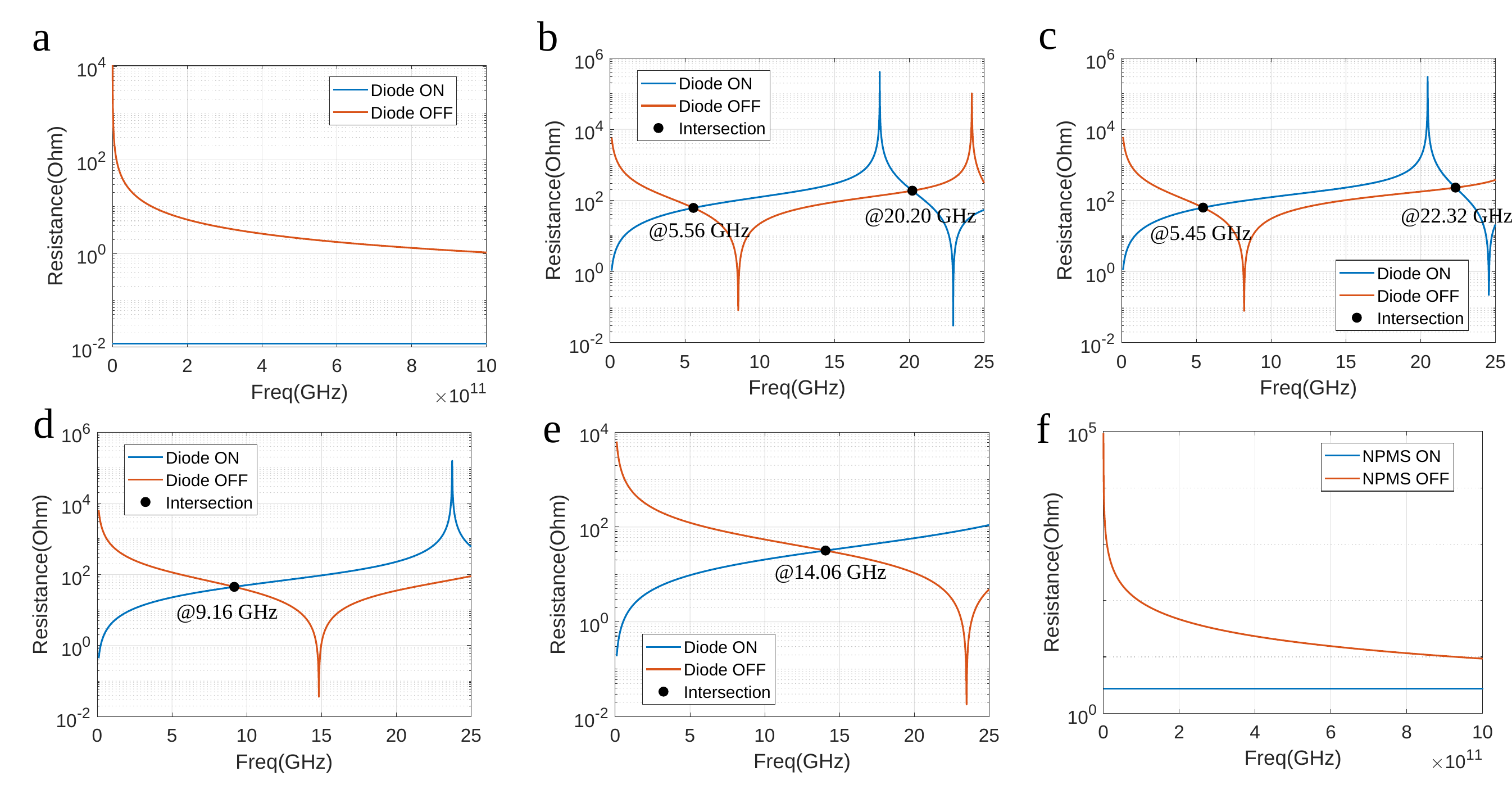}
  \caption{The intersection of ON-state and OFF-state impedance curves of SMP1320 diodes for \textbf{a,} Die package, \textbf{b,} SOT-23 package, \textbf{c,} SOD-323 package, \textbf{d,} SC-79 package,and \textbf{e,} SOD-882 package. \textbf{f,} The ON-state and OFF-state impedance curves of the NPMS.}
  \label{fig:diode-1}
\end{figure}

Fig. \ref{fig:diode-1} illustrates the intersection points of on-state and off-state impedance curves of SMP1320 diodes corresponding to different package types. The impedance curves of the diode with Die package exhibit no intersection point (Fig. \ref{fig:diode-1}a). At 100 GHz, the impedance difference between the ON and OFF states still reaches 10 $\Omega$. Nevertheless, owing to package parasitics, the impedance difference diminishes to zero (i.e., the intersection point) at a substantially lower frequency. At the intersection frequency, the diode loses its switching functionality. The intersection points of impedance curves occur at lower frequencies of 5.56 GHz (Fig. \ref{fig:diode-1}b) and 5.45 GHz (Fig. \ref{fig:diode-1}c) for the larger-sized SOT-23 and SOD-323 packages, respectively. This indicates that they can only be applied to frequency bands below the corresponding intersection frequencies. The smaller SC-79 (Fig. \ref{fig:diode-1}d) and SOD-882 (Fig. \ref{fig:diode-1}e) packages exhibit higher intersection frequencies of 9.16 GHz and 14.06 GHz, respectively, implying that the switch is applicable to higher operating frequency bands. However, there is also a conflict between compact, low-parasitic packages and high heat-dissipation capability. Consequently, in adaptive high-power electromagnetic protection scenarios that demand high frequency, high tolerance, and fast response, the diode exhibits clear limitations. By comparison, no intersection point exists on the impedance traces of the nanoplasma switch (NPMS) shown in Fig. \ref{fig:diode-1}f. Even at 400 GHz, a 20 $\Omega$ impedance difference is sustained between ON and OFF states, which reveals that the NPMS supports operation at frequencies substantially above the feasible frequency range of diodes.

The heat dissipation of NPMS devices and PIN diodes can be quantitatively described by a typical thermal time constant $\tau$, which is attributed to the time required for heat to diffuse through the entire switch and substrate to increase temperature. Following a thermal model that describes the heat equation in a circular thin plate\cite{dolleman2018transient}, the temperature of the entire system $T$ at the frequencies $f$ far below resonance is given by:
\begin{equation}
     T = \frac{T_{\text{th}}}{i2\pi f \tau +1}, 
   \label{eq:thermal delay equation}
\end{equation}
where $T_{\text{th}}$ is the temperature at the heat source. 

\begin{figure}[H]
  \centering
  \includegraphics[width=0.9\linewidth]{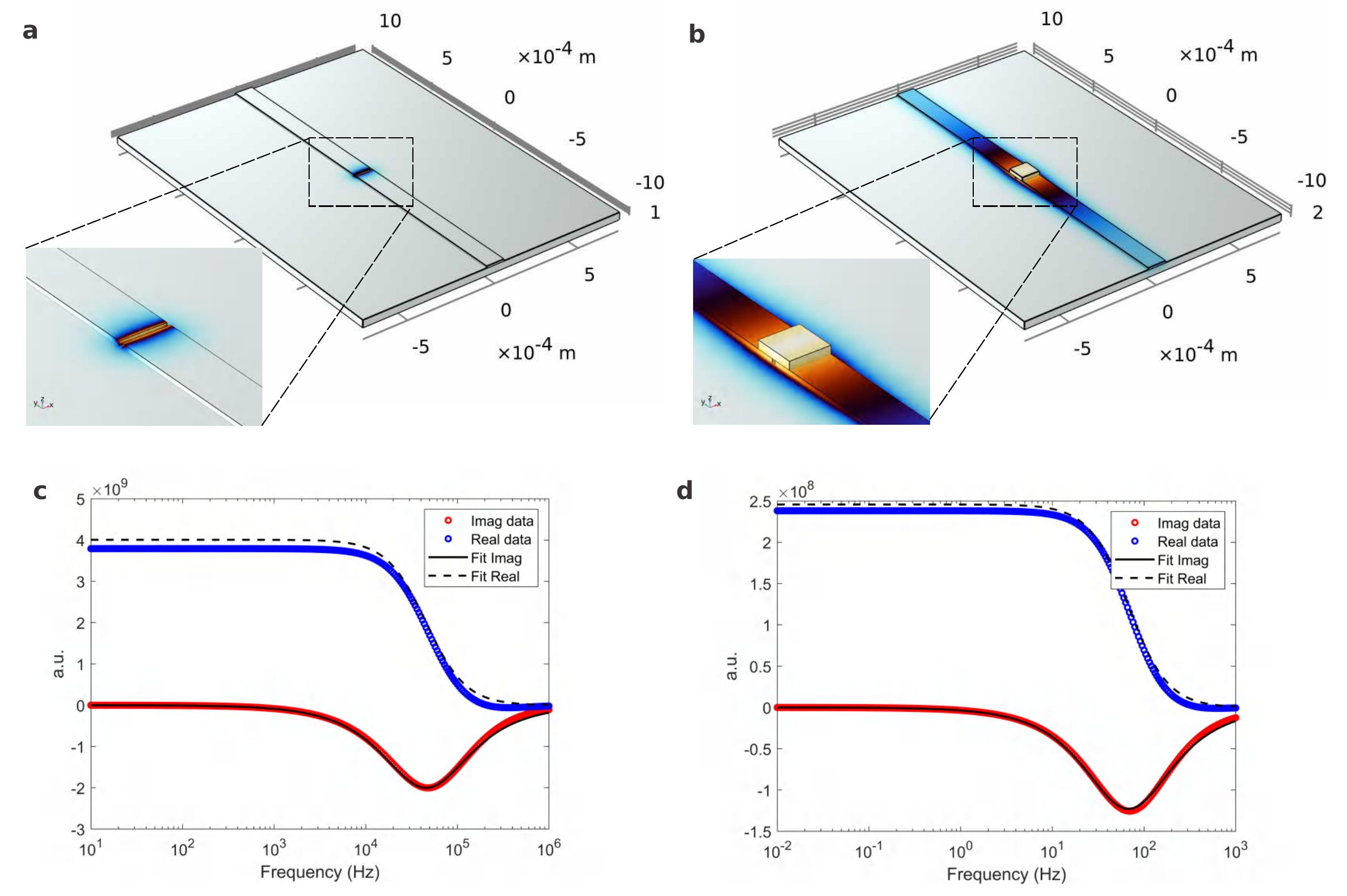}
  \caption{COMSOL simulations of thermal transport in NPMS and diode-based switch. \textbf{a} and  \textbf{b,} Schematic diagram and temperature distribution on COMSOL for NPMS and diode-based switch, respectively. \textbf{c} and \textbf{d,} Simulation results of average temperature $T_{\text{ave}}$ versus heating frequency $f$, which are fitted with Eq.~\ref{eq:thermal delay equation} to extract the thermal time constant $\tau$ of the NPMS and diode-based switch.}
  \label{fig:thermal_diode}
\end{figure}

The "heat transport in thin shells" module of COMSOL Multiphysics allows us to simulate the thermal transport in NPMS devices and diodes loaded on the substrate. We build a simple 3D domain as illustrated in Figs.~\ref{fig:thermal_diode}a and \ref{fig:thermal_diode}b, where heat sources are set on the NPMS nanogap and top surface of PIN, respectively. We define the thermal insulation condition  on the bottom of the substrate, with an initial temperature $T_0 = 273.15$~\si{K}. The mesh size is using "Extra Fine" to ensure the simulation accuracy of thermal transport. We then define a harmonic perturbation of heating with a frequency that varies from 10 to 10$^6$~\si{Hz} (NPMS) and 10$^{-2}$ to 10$^{3}$~\si{Hz} (diode), and thus to obtain the average temperature $T_{\text{ave}}$ of the membrane as a function of heating frequency $f$, as shown in Figs.~\ref{fig:thermal_diode}c and \ref{fig:thermal_diode}d. The heat capacity, mass density and thermal conductivity for NPMS devices (SiC substrate and GaN electrode materials as used in our work) and diodes (FR4 substrate, Au electrode, GaAs and Si diode materials as commonly used before) are substituted into the model. By fitting Eq.~\ref{eq:thermal delay equation} to the simulated $T_{\text{ave}}$ versus $f$, we extract the thermal time constant $\tau$ about $2.366\times 10^{-3}$ and $2.554\times 10^{-3}$~\si{s} for GaAs- and Si-based diodes, respectively, and $3.489\times 10^{-6}$~\si{s} for NPMS devices. Therefore, we confirm a much enhanced heat dissipation in the proposed on-chip nanoplasma compared with that in conventional devices.

\section{Local short-circuiting failure of all-metal nanoplasma switch}

\begin{figure}[H]
  \centering
  \includegraphics[width=0.6\linewidth]{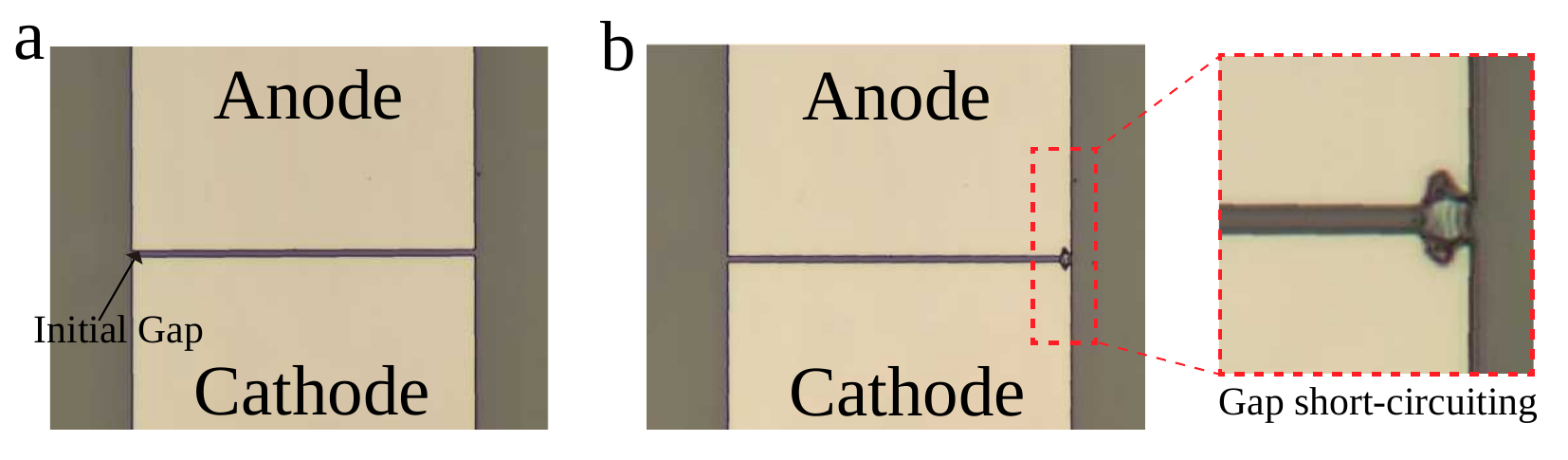}
  \caption{\textbf{a,} Initial state of the metal electrode gap. \textbf{b,} Molten-metal-induced gap short circuit. }
  \label{fig:Pt_shorted-circuit}
\end{figure}

The NPMS is capable of carrying ampere-level on-state currents and has a high power-handling capacity. Nevertheless, this imposes extremely high current density and power density loads on the switch electrodes. For NPMS devices equipped with metal electrodes, localized electrode temperatures under the on-state operation can reach the melting point of the metal, triggering melting of the metal over a certain area. Once the molten metal migrates into the nanoscale electrode gap, there is an extremely high risk of gap short-circuiting and subsequent switch failure. As an example, Fig.~\ref{fig:Pt_shorted-circuit} presents the molten metal-induced gap short-circuit of an all-metal NPMS. The metal electrode consists of a Ti adhesion layer and a Pt functional layer with thicknesses of 20 nm and 500 nm, respectively. 

\section{Preparation of the GaN-based nanoplasma switch}
\begin{figure}[H]
  \centering
  \includegraphics[width=0.7\linewidth]{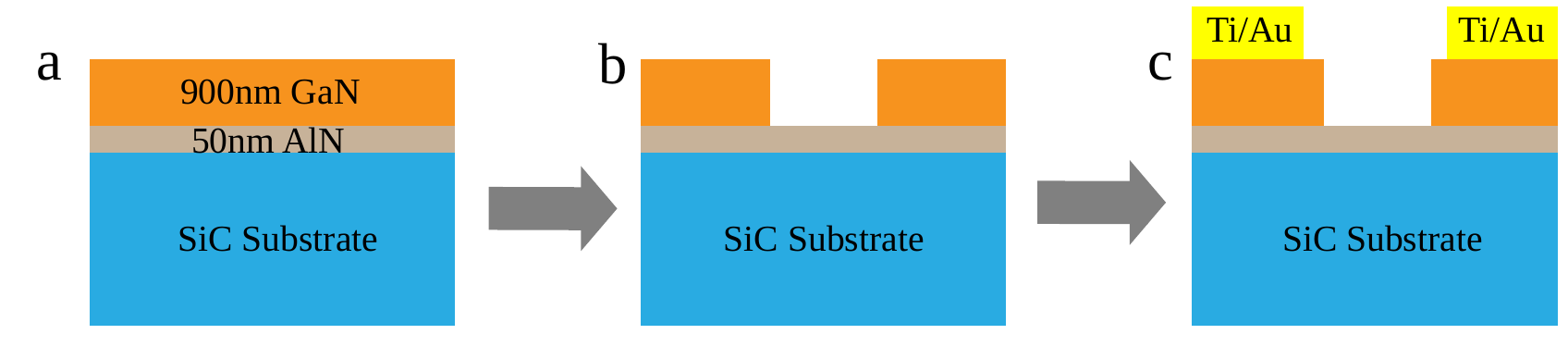}
  \caption{Schematic of the fabrication process. \textbf{a,} Layer schematic of GaN epitaxial wafer. \textbf{b,} Cross-sectional schematic of the GaN electrode gap. \textbf{c,} Schematic of the metal contact electrodes on GaN.}
  \label{fig:fabrication}
\end{figure}

A 900 nm thick GaN film with a 50 nm AlN buffer layer was epitaxially grown using MOCVD on semi-insulating SiC substrates (Fig.~\ref{fig:fabrication}a).  The GaN layer is n-type doped with a doping concentration of \(3\times10^{18}\ \mathrm{cm^{-3}}\), and the resulting GaN film achieves a sheet resistance of \(47\ \Omega/\mathrm{sq}\). 

The NPMS was fabricated using a two-step photolithography process. The first photolithography and ion-beam etching steps were used to define the GaN structures (Fig.~\ref{fig:fabrication}b), whereas the second photolithography, metal deposition, and lift-off steps were employed to form the Ti/Au (20 nm/100 nm) electrodes (Fig.~\ref{fig:fabrication}c). The detailed fabrication procedure is described below.

\subsection*{Patterning and ion-beam etching of GaN}

\begin{enumerate}
    \item \textbf{Surface pretreatment and photoresist coating.}
    The wafer was first treated in an HMDS system to improve the adhesion between the substrate and the photoresist. AR80 photoresist was then dispensed onto the sample and spin-coated at \(500~\mathrm{rpm}\) for \(5~\mathrm{s}\), followed by \(3000~\mathrm{rpm}\) for \(30~\mathrm{s}\). The coated sample was soft-baked at \(95\,^{\circ}\mathrm{C}\) for  \(90~\mathrm{s}\).

    \item \textbf{First photolithographic exposure.}
    The desired pattern was transferred into the photoresist using a Nikon I12 stepper with an exposure time of \(300~\mathrm{ms}\).

    \item \textbf{Development and inspection.}
    The exposed photoresist was developed in a \(2.38\%\) TMAH aqueous developer for \(45~\mathrm{s}\). The developed pattern was subsequently inspected using a high-magnification optical microscope to verify the pattern fidelity and identify possible lithographic defects.

    \item \textbf{Photoresist hard bake.}
    After development, the sample was baked at \(100\,^{\circ}\mathrm{C}\) for \(2~\mathrm{min}\). This step hardened
    the photoresist and removed residual solvent and moisture, thereby improving the resistance of the photoresist mask during the subsequent etching process.

    \item \textbf{Oxygen-plasma descum.}
    Residual photoresist at the bottom of the developed openings was removed by oxygen-plasma cleaning using an M4L system. The plasma power and treatment duration were \(200~\mathrm{W}\) and \(2~\mathrm{min}\), respectively.

    \item \textbf{Ion-beam etching.}
    The exposed \(\mathrm{GaN}\) layer was etched using an ion-beam etching (IBE) system. The ion energy, beam current, and etching time were  \(400~\mathrm{eV}\), \(100~\mathrm{mA}\), and \(90~\mathrm{min}\), respectively. The patterned photoresist served as the etching mask, allowing the unprotected \(\mathrm{GaN}\) regions to be removed.

    \item \textbf{Photoresist removal.}
    Following ion-beam etching, the remaining photoresist was removed by ultrasonic cleaning in acetone and isopropyl alcohol. The sample was subsequently rinsed with deionized water for \(2~\mathrm{min}\) and dried using a nitrogen gun.
\end{enumerate}

\subsection*{Fabrication of the Ti/Au electrodes}

\begin{enumerate}
    \item \textbf{Surface pretreatment and second photoresist coating.}
    The etched sample was treated again in an HMDS system. A layer of 1303 photoresist was then spin-coated at \(500~\mathrm{rpm}\) for \(5~\mathrm{s}\), followed by \(3000~\mathrm{rpm}\) for \(30~\mathrm{s}\). The sample was soft-baked at \(110\,^{\circ}\mathrm{C}\) for \(120~\mathrm{s}\).

    \item \textbf{Second photolithographic exposure.}
    The electrode pattern was defined using an MA6 mask aligner with an exposure time of \(4~\mathrm{s}\). After exposure, the sample was baked at \(110\,^{\circ}\mathrm{C}\) for \(2~\mathrm{min}\).

    \item \textbf{Development and inspection.}
    The exposed photoresist was developed in a \(2.38\%\) TMAH aqueous developer for \(40~\mathrm{s}\). The resulting pattern was inspected using a high-magnification optical microscope before metal deposition.

    \item \textbf{Ti/Au deposition.}
    A Ti/Au metal stack was deposited using a magnetron sputtering system. The thicknesses of the Ti adhesion layer and the Au electrode layer were \(20~\mathrm{nm}\) and \(100~\mathrm{nm}\), respectively.

    \item \textbf{Lift-off.}
    The sample was immersed in acetone and subsequently ultrasonically  cleaned in acetone and isopropyl alcohol to remove the photoresist and the metal deposited on top of it. The sample was then rinsed with deionized water and dried using a nitrogen gun, leaving the patterned Ti/Au electrodes on the substrate.

    \item \textbf{Device dicing.}
    Finally, the fabricated wafer was diced along the predefined alignment  marks using a DISCO dicing system to obtain individual micro/nanoscale plasma-switch devices.
\end{enumerate}

In summary, the fabrication process consisted of HMDS surface treatment, photoresist coating, photolithographic pattern transfer, oxygen-plasma descumming, ion-beam etching of the \(\mathrm{GaN}\) layer, a second photolithography process, Ti/Au magnetron sputtering, metal lift-off, and device dicing.

\begin{figure}[H]
  \centering
  \includegraphics[width=0.8\linewidth]{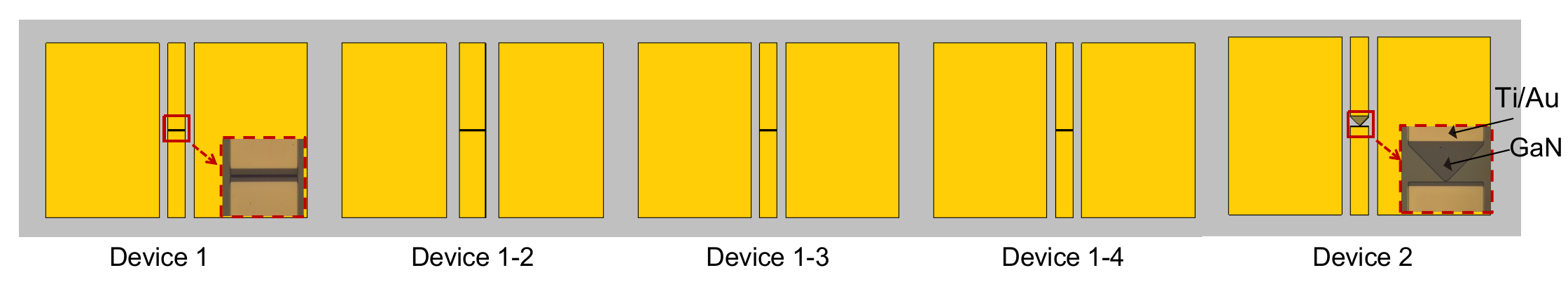}
  \caption{NPMS devices with different dimensions and structures.}
  \label{fig:device_structure}
\end{figure}

Fig.~\ref{fig:device_structure} illustrates five NPMS devices with different dimensions and structures. Devices 1, 1-2, 1-3, and 1-4 feature parallel edge-to-edge electrode structures. Device 1 has an electrode gap of 800 nm and an electrode width of 100 \si{\micro\meter}. Device 1-2 has an electrode gap of 800 nm and an electrode width of 150 \si{\micro\meter}. Device 1-3 has an electrode gap of 1000 nm and an electrode width of 100 \si{\micro\meter}. Device 1-4 has an electrode gap of 2000 nm and an electrode width of 100 \si{\micro\meter}. Device 2 features a tip-to-edge electrode structure with an electrode gap of 800 nm and an electrode width of 100 \si{\micro\meter}.

\section{Conduction response simulation of the nanoplasma switch}
The NPMS turns on and off through the formation and disappearance of plasma inside its nanoscale electrodes. When a voltage exceeding the threshold is applied, gas discharge occurs in the gap, forming a highly conductive plasma that turns on the NPMS. In general, the gas discharge process can be quantitatively described by a coupled system of transport equations for charged particles, centered on a fully coupled solution framework consisting of the continuity equation~\eqref{1}, poisson equation~\eqref{2}, and charge-density relation \eqref{3}\cite{ZhangNing-129}:
\begin{equation}
    \frac{\partial n_i}{\partial t} + \nabla \cdot \left(w_i n_i - D_i \nabla n_i\right)= R_i \\
    \label{1}
\end{equation}

\begin{equation}
    \nabla \cdot \left(\varepsilon_r \varepsilon_0 \mathbf{E}\right)= \rho\\
    \label{2}
\end{equation}

\begin{equation}
    \rho=e \sum_i z_i n_i 
    \label{3}
\end{equation}
where the subscripts \(i=e,p,n\) denote electrons, positive ions and negative ions, respectively; \(n_i\) is the number density of charged particles, \(\mathbf{E}\) is the electric field, \(z_i\) is the charge number of carriers, \(w_i\) is the drift velocity in the electric field, \(D_i\) is the diffusion coefficient, \(R_i\) is the reaction rate and \(e\) is the elementary charge. Equations~\eqref{1} \textendash~\eqref{3} describe the drift–diffusion–reaction transport of charged particles and the modulation of the electric field by space charge. To initiate and sustain the discharge process, the cathode must continuously supply seed electrons.

The turn-on triggering mechanism of NPMS has been primarily interpreted using microscale gas-breakdown models derived from the modified Paschen framework, in which cathode field emission (FE) supplies seed electrons and facilitates gap breakdown\cite{LovelessGarner-108}. The FE process is classically described by Fowler-Nordheim (FN) theory\cite{Wolfram,NguyenKang-124}, giving the FE current density $J_{\mathrm{FE}}$ as below:
\begin{equation}
    J_{\mathrm{FE}} = A_{\mathrm{FN}} \frac{F^{2}}{\phi} \exp \left( -\frac{B_{\mathrm{FN}} \phi^{3 / 2}}{F} \right)
   \label{eq:Field emission}
\end{equation} 
where $\phi$ (unit: eV) is the work function of cathode material,  $F$ (unit: V/m) is the local electric field at the cathode, $A_{\mathrm{FN}}$ and $B_{\mathrm{FN}}$ are constants equal to \(1.54 \times 10^{-6}\,\mathrm{A\,V^{-2}\,eV}\) and \(6.83 \times 10^{9}\,\mathrm{eV^{-3/2}\,m^{-1}\,V}\), respectively. The cathode local electric field $F$ is typically enhanced from the applied field $E$. The field enhancement effect is generally described by a field enhancement factor $\beta$ ($F= \beta E$), which represents the geometric effect of the electrode surface, usually depending on the local roughness and defects. 

\begin{figure}[H]
  \centering
  \includegraphics[width=0.6\linewidth]{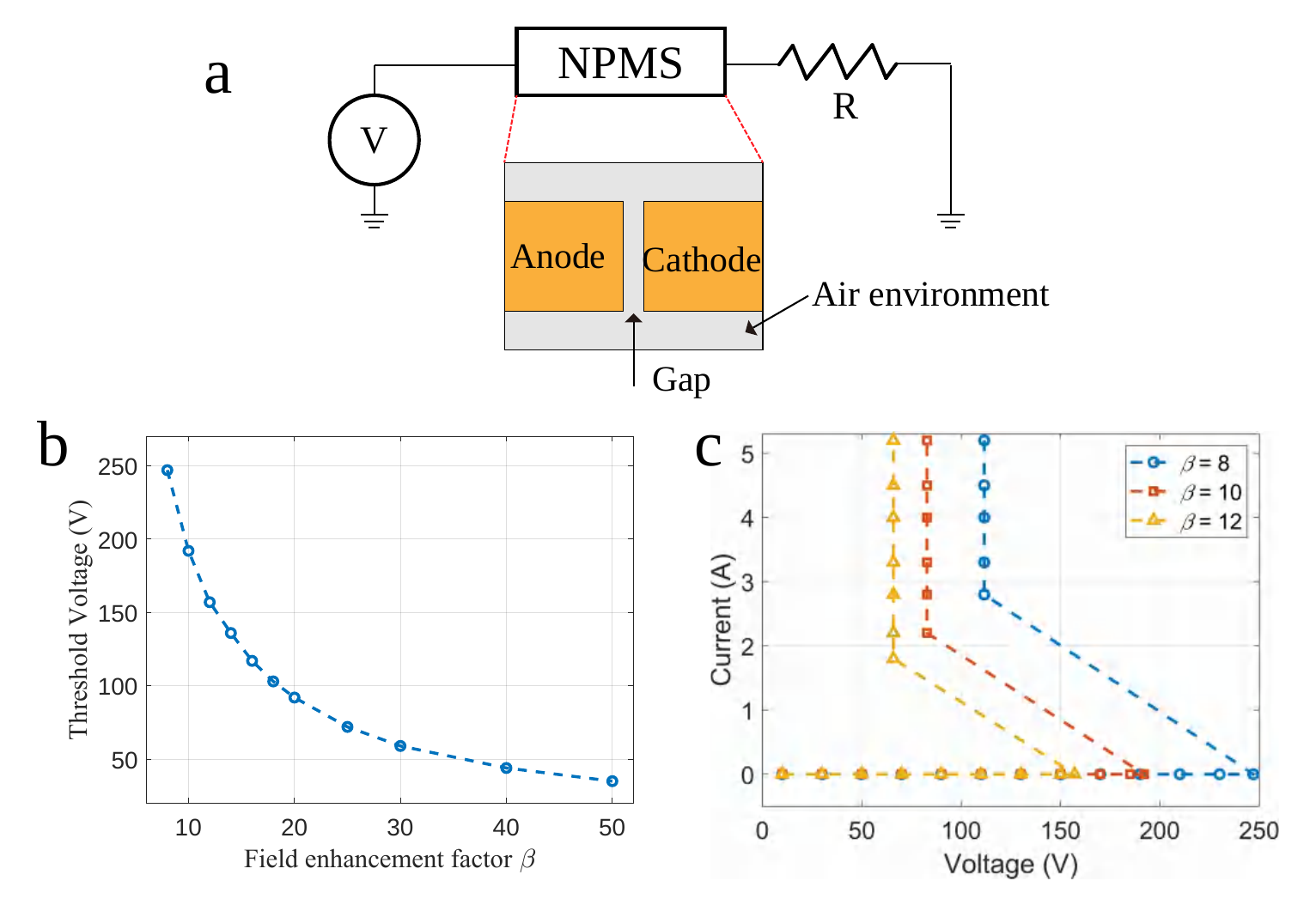}
  \caption{Simulation of the NPMS conduction response. \textbf{a,} Complete external circuit model diagram. \textbf{b,} Threshold voltage versus the field enhancement factor $\beta$. \textbf{c,} Simulated current-voltatge curves corresponding to $\beta$ = 8, 10, and 12.}
  \label{fig:simulation_model}
\end{figure}

We employ the "electrical discharge" and "circuit" module of COMSOL Multiphysics to establish a gas discharge numerical model driven by the FE mechanism. Fig.~\ref{fig:simulation_model}a illustrates the complete external circuit model diagram, consisting of a voltage source, a current-limiting resistor of 50 $\Omega$, and an NPMS. The NPMS is simplified to a two-dimensional model in an air environment. The high electric field in the nanoscale gap enhances the FE effect, which is essential for seed electron generation. The electric field across the entire gap accelerates the electrons and ions to high energy, leading to the gas breakdown. The air domain bounded by the electrodes constitutes the computational region of the model. The FE process of the cathode is incorporated into the simulations by converting the FE current density $J_{\mathrm{FE}}$ into the electron emission flux $\boldsymbol{\Gamma}_\mathrm{e}$ at the cathode boundary using Eq. \eqref{eq:Flux}. Here, $\mathbf{n}$ denotes the unit normal vector at the cathode interface pointing from the gap to the cathode, and $\mathrm{e}$ represents the elementary charge.
\begin{equation}
    -\mathbf{n} \cdot \boldsymbol{\Gamma}_\mathrm{e} = \frac{J_{\mathrm{FE}}}{e}
   \label{eq:Flux}
\end{equation} 

With the structural parameters and electrode material fixed, the conduction response characteristics of the NPMS devices are predominantly governed by the field enhancement factor $\beta$. For example, we set the gap distance $d$ between electrodes to 800 nm, the width $w$ of the electrode is 100 $\mu$m, the electrode thickness is 900 nm, and the work function of the electrode material is 4 eV\cite{WeiChen-305}. The simulation results show that an increase in the field enhancement factor $\beta$ leads to a reduction in the switching threshold voltage (Fig. \ref{fig:simulation_model}b and c).

\begin{figure}[H]
  \centering
  \includegraphics[width=\linewidth]{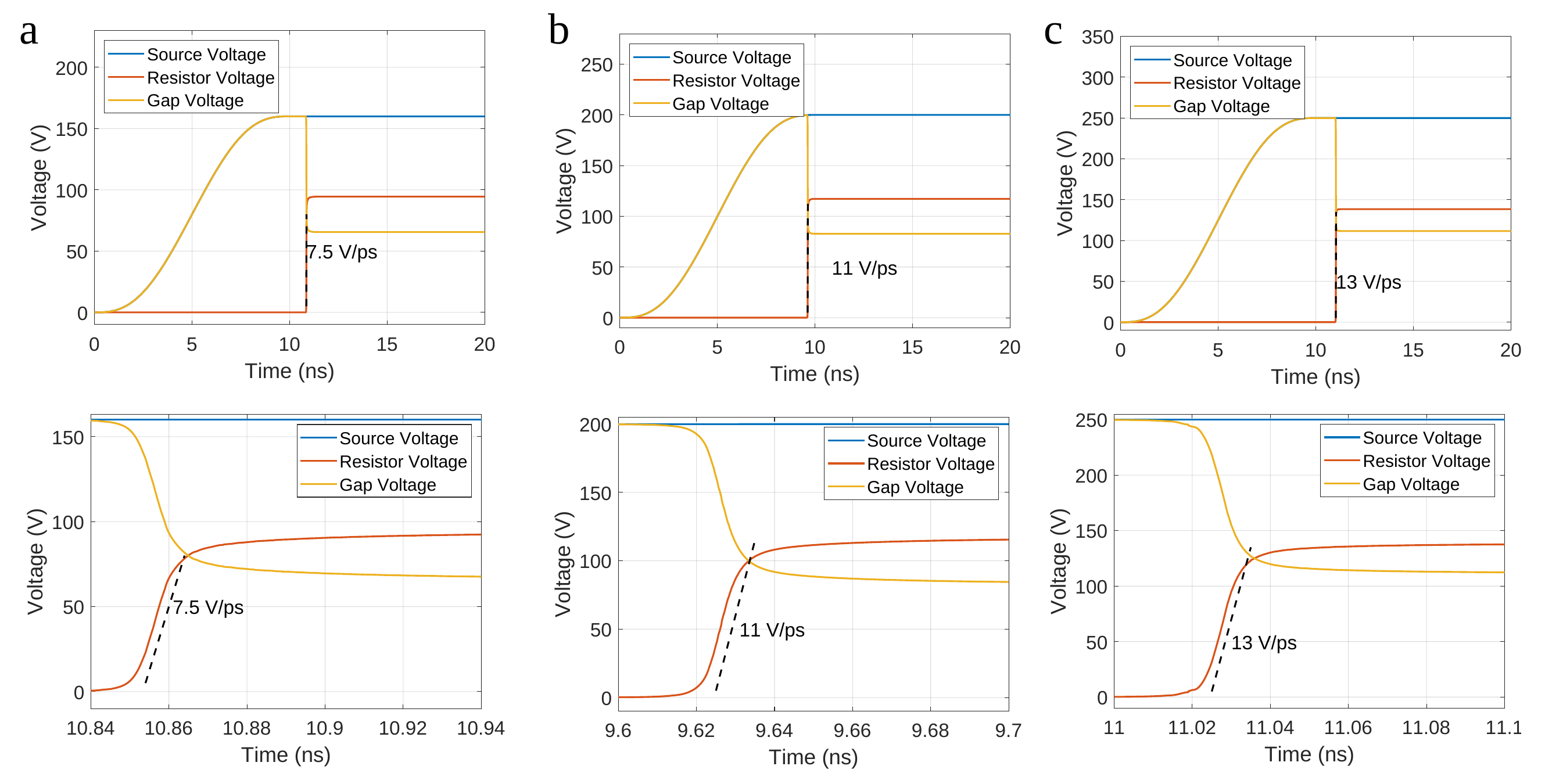}
  \caption{Simulated transient voltage responses at threshold voltages of \textbf{a,} 157 V, \textbf{b,} 192 V, and \textbf{c,} 247 V.}
  \label{fig:IV_response_time}
\end{figure}

We can see from Fig. \ref{fig:IV_response_time} that despite differences in the switch threshold voltage, the turn-on transition time remains nearly identical (around 15 ps). This indicates that a higher threshold voltage corresponds to a faster turn-on speed of the switch. In detail, the turn-on speeds of 7.5 V/ps, 11 V/ps and 13 V/ps are associated with threshold voltages of 157 V, 192 V and 247 V in sequence.

\section{Measurement of  current-voltage characteristics of the nanoplasma switch}

The current-voltage (\textit{I-V}) characteristics of the five fabricated NPMS devices are measured using the standard transmission-line pulse (TLP) method. A HPPI-TLP-4010C pulse generator was employed to apply voltage pulses to the device under test, while the corresponding transient voltage and current waveforms were recorded using a Tektronix MOS64 digital oscilloscope with a sampling rate of 25 GS/s and a bandwidth of 1 GHz. The applied pulses had a width of \(100~\mathrm{ns}\) and a rise time of \(10~\mathrm{ns}\). The pulsed measurement minimized the influence of long-term Joule heating and enabled characterization of the transient electrical response of the device. For each pulse, the voltage and current values are extracted from the flat-top region of the recorded waveforms and averaged to obtain a single data point on the \textit{I-V} curve. The pulse amplitude is then increased in set steps, and the measurement sequence is repeated to obtain the desired \textit{I-V} curve.

\begin{figure}[H]
  \centering
  \includegraphics[width=0.6\linewidth]{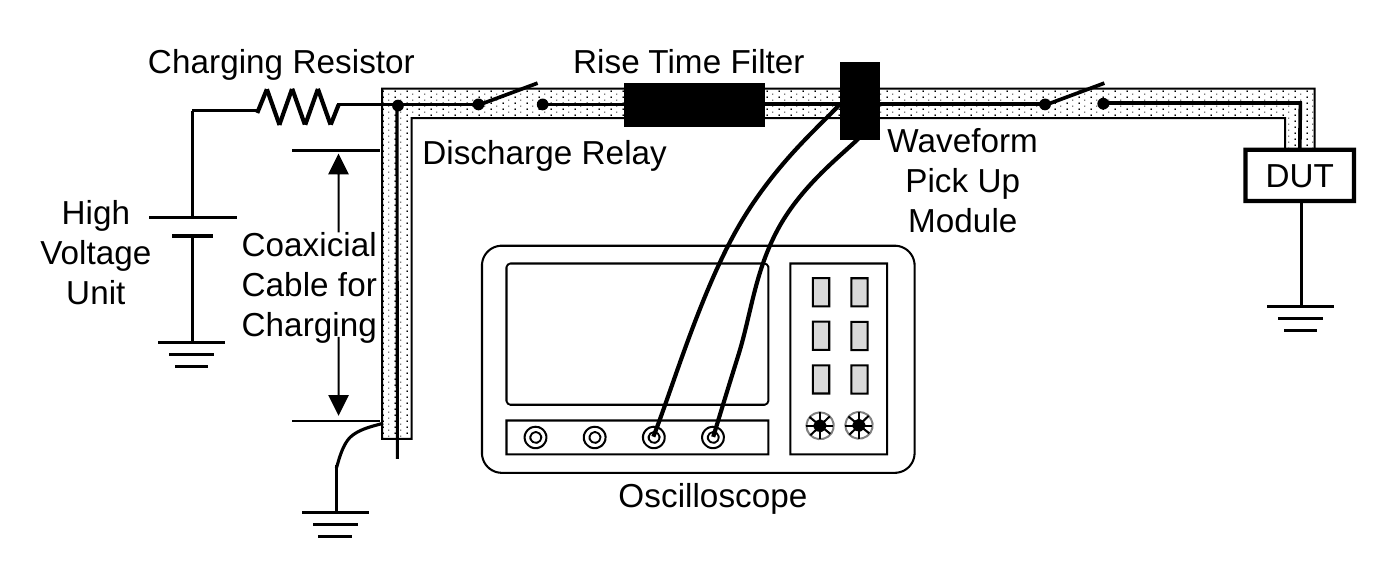}
  \caption{TLP test setup diagram.}
  \label{fig:TLP}
\end{figure}

\begin{figure}[H]
  \centering
  \includegraphics[width=\linewidth]{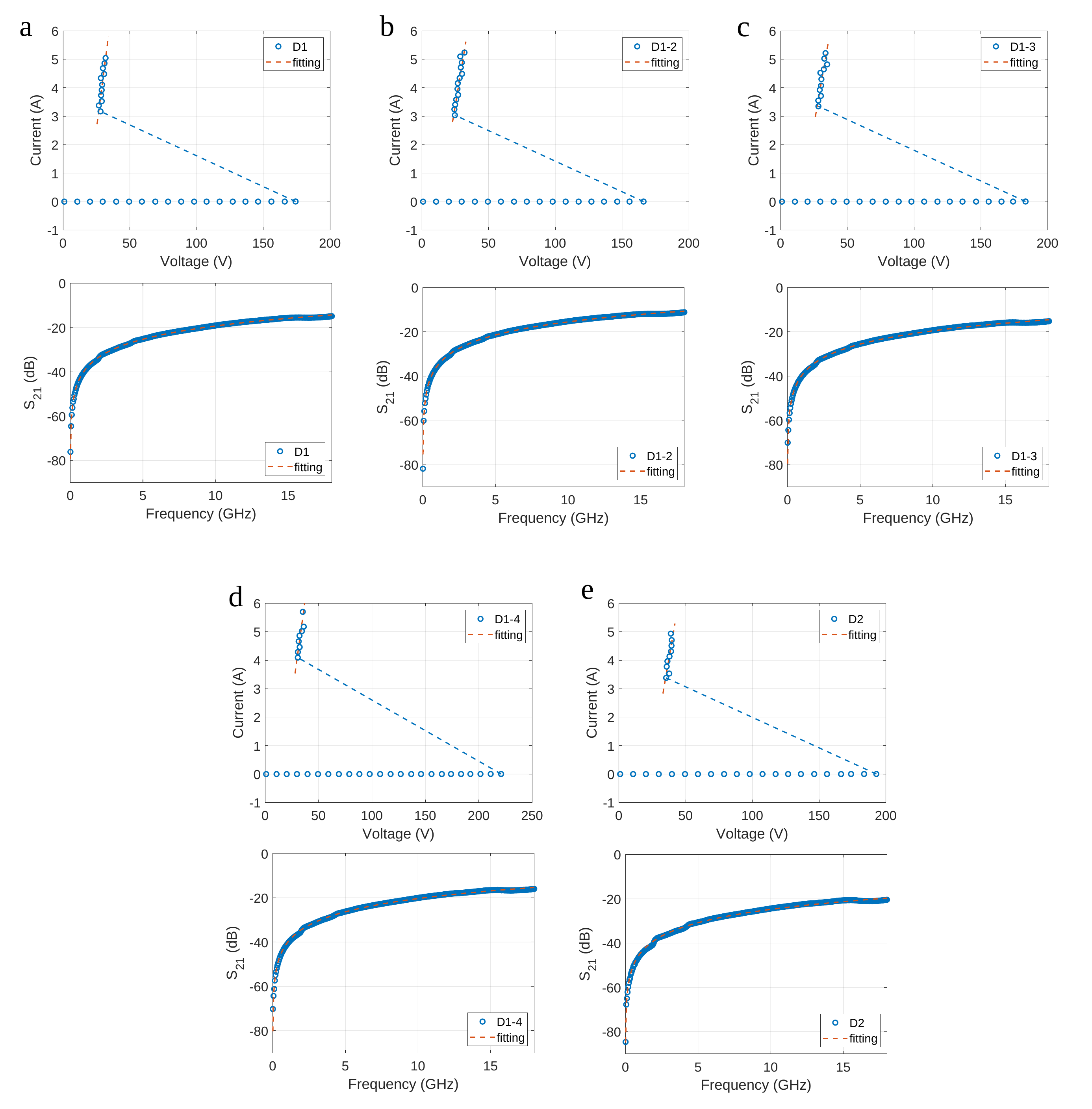}
  \caption{\textit{I-V} curves and transmission coefficients of NPMS devices. Subfigures \textbf{a-e} correspond to devices 1, 1-2, 1-3, 1-4, and 2, respectively.}
  \label{fig:IV_S21}
\end{figure}

\begin{figure}[H]
  \centering
  \includegraphics[width=0.4\linewidth]{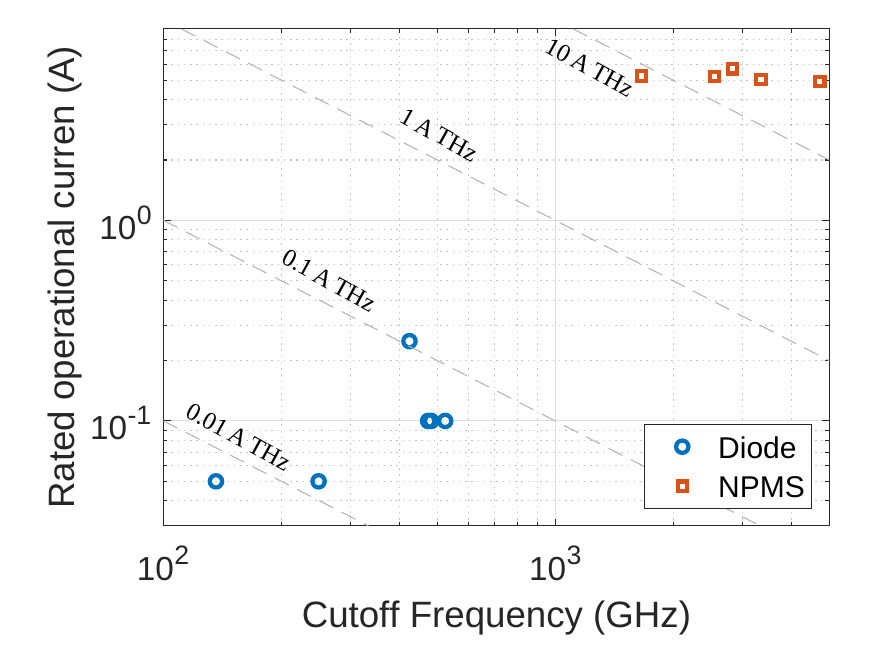}
  \caption{Rated operational current versus cut-off frequency for diodes and NPMS devices.}
  \label{fig:switchFOM}
\end{figure}

Fig. \ref{fig:IV_S21} shows the \textit{I-V} curves and off-state transmission coefficients of NPMS devices illustrated in Fig. \ref{fig:device_structure}. As can be seen, these devices demonstrate high on-state currents exceeding 3~A and low on-state dynamic resistances $R_{\mathrm{ON}}$ lower than 5 $\Omega$. They also exhibit small transmission coefficients even in the Ku band, indicating their small cut-off capacitances $C_{\mathrm{OFF}}$.  The corresponding fitting $R_{\mathrm{ON}}$ and $C_{\mathrm{OFF}}$  are listed in Table \ref{tab:dynamic_resistance_capacitance}. Table \ref{tab:commercial_diodes} lists the $R_{\mathrm{ON}}$ and $C_{\mathrm{OFF}}$ of commercial diodes according to their official datasheets. Their cut-off frequencies and absolute maximum currents (termed  rated operational current in the main text) are also provided in the Tables. The measured NPMS devices exhibit a rated operational current versus cut‑off frequency product ($I_{\text{ro}} \times f_{c,}$) that is nearly two orders of magnitude higher than that of commercial diodes (Fig. \ref{fig:switchFOM}).

\begin{table}[htbp]
\centering
\caption{Electrical characteristics of NPMS devices.}
\label{tab:dynamic_resistance_capacitance}
\renewcommand{\arraystretch}{1.5}
\begin{threeparttable}
\scriptsize
\setlength{\tabcolsep}{5pt} 
\begin{tabular}{l|cccc}
\hline
 & On-state resistance & Cut-off capacitance &
 Cut-off Frequency (GHz) & Current (A) \\
 & $R_{\mathrm{on}}$ ($\Omega$) & $C_{\mathrm{OFF}}$ (fF) & & \\
\hline
D1   & 2.8 & 17.3 & 3349.2 & 5.1 \\
D1-2 & 3.5 & 27.1 & 1660.8 & 5.2 \\
D1-3 & 3.7 & 16.8 & 2551.9 & 5.2 \\
D1-4 & 3.7 & 15.3 & 2830.7 & 5.7 \\
D2   & 3.7 & 9.2  & 4731.8 & 4.9 \\
\hline
\end{tabular}
Note: $R_{\mathrm{on}}$ and $C_{\mathrm{OFF}}$ are obtained by fitting the \textit{I--V} curves and transmission coefficients of the NPMS devices. The cut-off frequency is calculated as $1/(2\pi R_{\mathrm{ON}}C_{\mathrm{OFF}})$. The current corresponds to the maximum measured point on the \textit{I--V} curve.
\end{threeparttable}
\end{table}

\begin{table}[htbp]
\centering
\caption{Electrical characteristics of the commercial diodes.}
\label{tab:commercial_diodes}
\renewcommand{\arraystretch}{1.5}
\begin{threeparttable}
\scriptsize
\setlength{\tabcolsep}{5pt} 
\begin{tabular}{l|cccc}
\hline
& On-state resistance
& Cut-off capacitance
& Cut-off Frequency (GHz)
& Current (mA) \\
& $R_{\mathrm{on}}$ ($\Omega$)
& $C_{\mathrm{OFF}}$ (fF)
& & \\
\hline
BAP51-02      & 1.5 & 400& 248.8 &50 \\   
NSR201MX-D    & 14.0  & 150 &63.2&50 \\ 
MA4AGFCP910   & 5.2 & 18 & 637.6& /\\ 
SMP1345       & 1.5 & 190 & 482.3&100\\ 
MMP4401       & 0.8 & 350 &523.5&100\\ 
MA4L011-134   & 2.1 & 130& 473.7 &100 \\ 
NSVP249SDSF3  & 4.5 & 230 & 136.1 &50\\ 
MA4SPS402     & 5.0   & 45  &424.4&250 \\ 
\hline
\end{tabular}
      Note: $R_{\mathrm{on}}$ and $C_{\mathrm{OFF}}$
      are provided in the commercial-diode datasheets.
      The cut-off frequency is calculated as
      $1/(2\pi R_{\mathrm{on}}C_{\mathrm{OFF}})$, with an additional
      general parasitic capacitance of 30 fF taken into account.
      The reported current corresponds to the absolute maximum current
      specified in the datasheets.
  \end{threeparttable}
\end{table}

\section{Structural parameters of the proposed designs}
\subsection*{SPPM}
\begin{figure}[H]
  \centering
  \includegraphics[width=0.5\linewidth,angle=270]{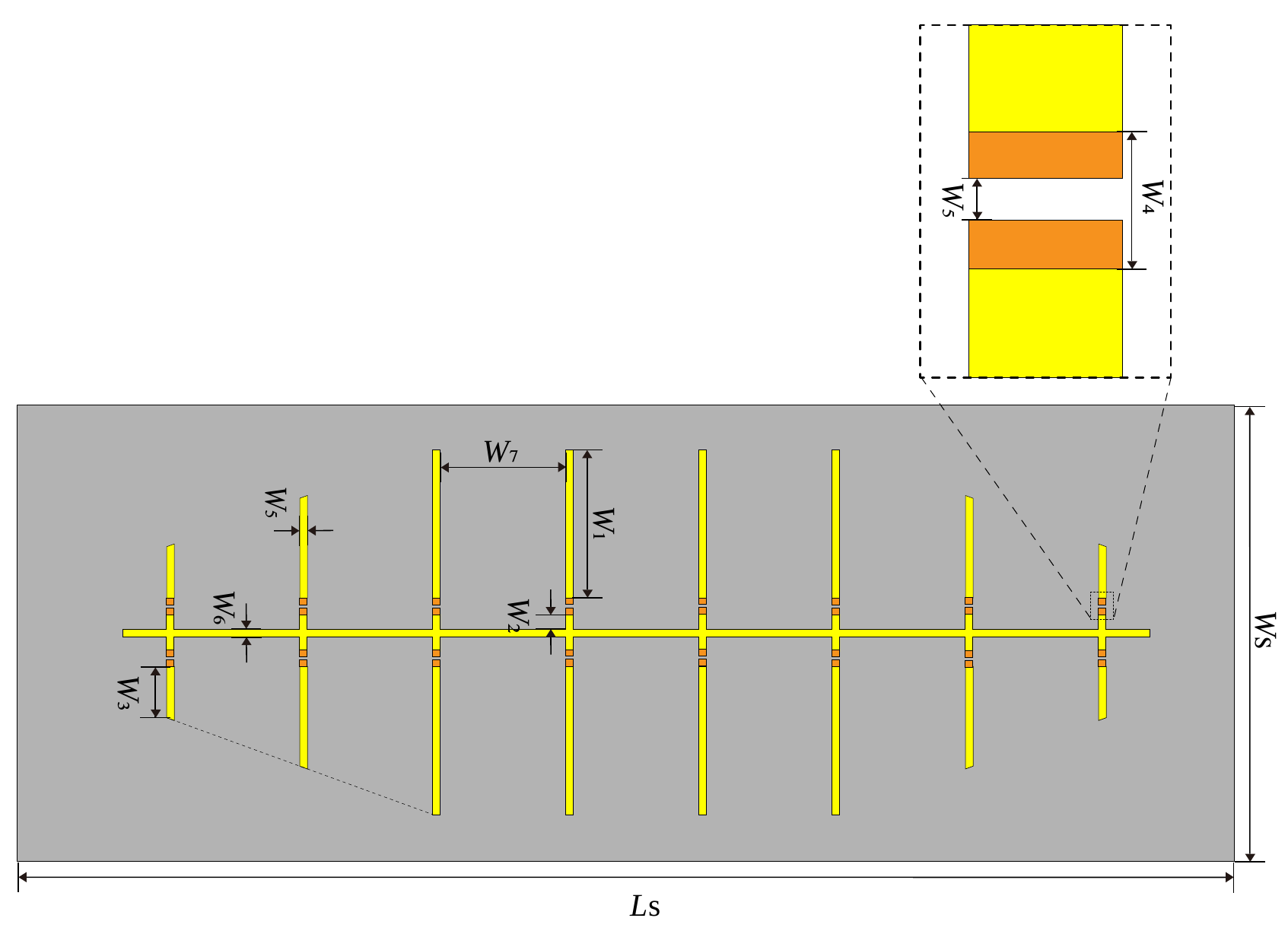}
  \caption{Structural parameters of the proposed SPPM-1.}
  \label{fig:SSPP}
\end{figure}

\begin{figure}[H]
  \centering
  \includegraphics[width=0.5\linewidth,angle=270]{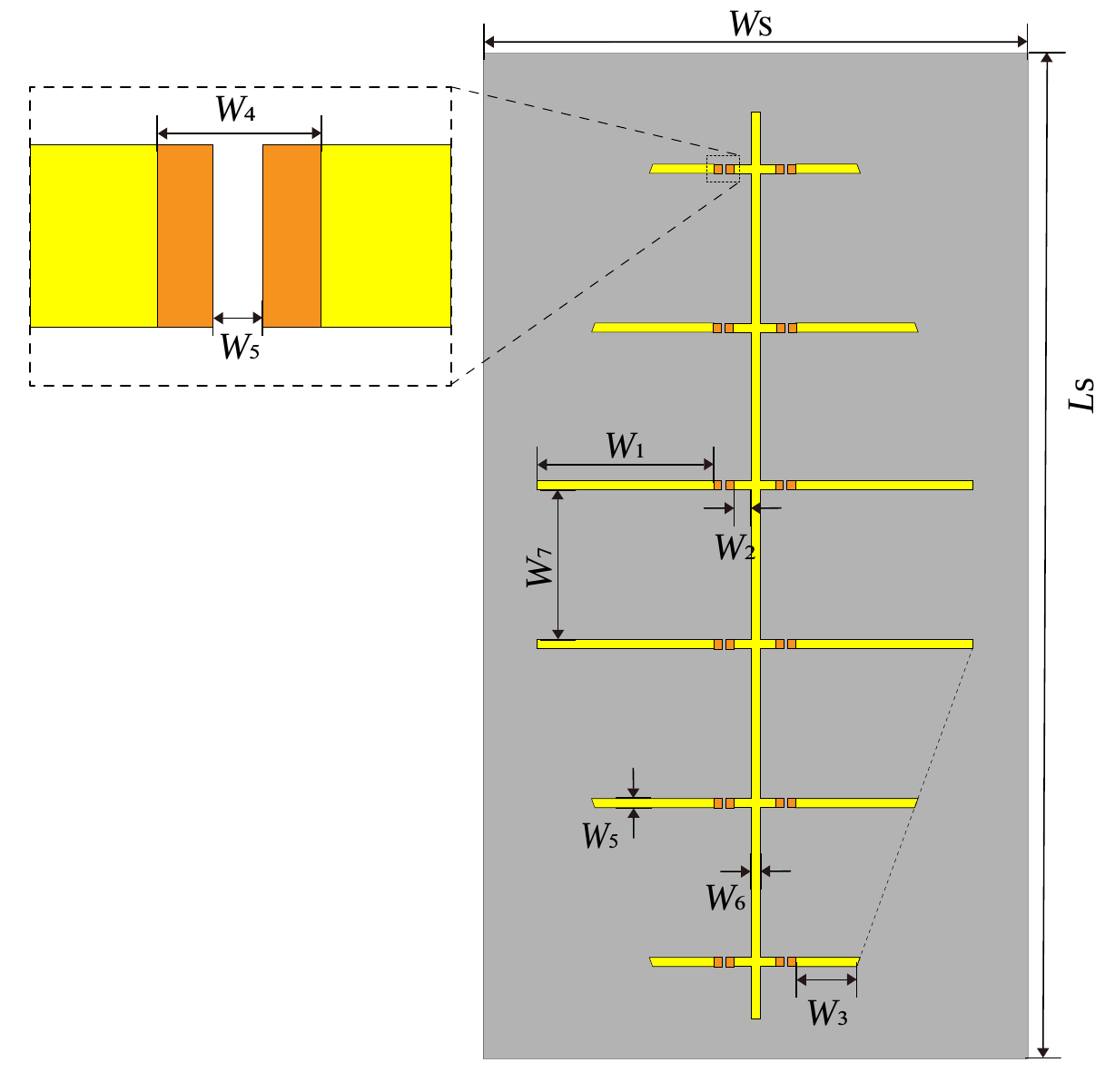}
  \caption{Structural parameters of the proposed SPPM-2.}
  \label{fig:SSPP-2}
\end{figure}

We design two SPPMs with similar structures, as shown in Fig. \ref{fig:SSPP} and Fig. \ref{fig:SSPP-2} (termed SPPM-1 and SPPM-2). Their detailed structural parameters are listed in Table \ref{tab:SPPM_params} and  Table \ref{tab:SPPM_params-2}, respectively.

\begin{table}[htbp]
\centering
\caption{Detailed structural parameters of the designed SPPM-1}
\label{tab:SPPM_params}
\renewcommand{\arraystretch}{1.2}
\scriptsize
\setlength{\tabcolsep}{5pt} 
\begin{tabular}{@{}lccccccccc@{}}
\hline
\textbf{Param.} & $L_s$ & $W_s$ & $W_1$ & $W_2$ & $W_3$ & $W_4$  & $W_5$  & $W_6$  & $W_7$ \\
\textbf{Value}  & 16 mm & 6 mm & 1.95 mm & 0.19 mm & 0.67 mm & 10 \si{\micro\meter}  &800 nm & 0.1 mm & 1.65 mm\\
\hline
\end{tabular}
\end{table}

\begin{table}[htbp]
\centering
\caption{Detailed structural parameters of the designed SPPM-2}
\label{tab:SPPM_params-2}
\renewcommand{\arraystretch}{1.2}
\scriptsize
\setlength{\tabcolsep}{5pt} 
\begin{tabular}{@{}lccccccccc@{}}
\hline
\textbf{Param.} & $L_s$ & $W_s$ & $W_1$ & $W_2$ & $W_3$ & $W_4$  & $W_5$  & $W_6$  & $W_7$ \\
\textbf{Value}  & 10 mm & 7.8 mm & 3.4 mm & 0.19 mm & 2 mm & 10 \si{\micro\meter}  &800 nm & 0.1 mm & 1.65 mm\\
\hline
\end{tabular}
\end{table}

\begin{figure}[H]
  \centering
  \includegraphics[width=0.3\linewidth]{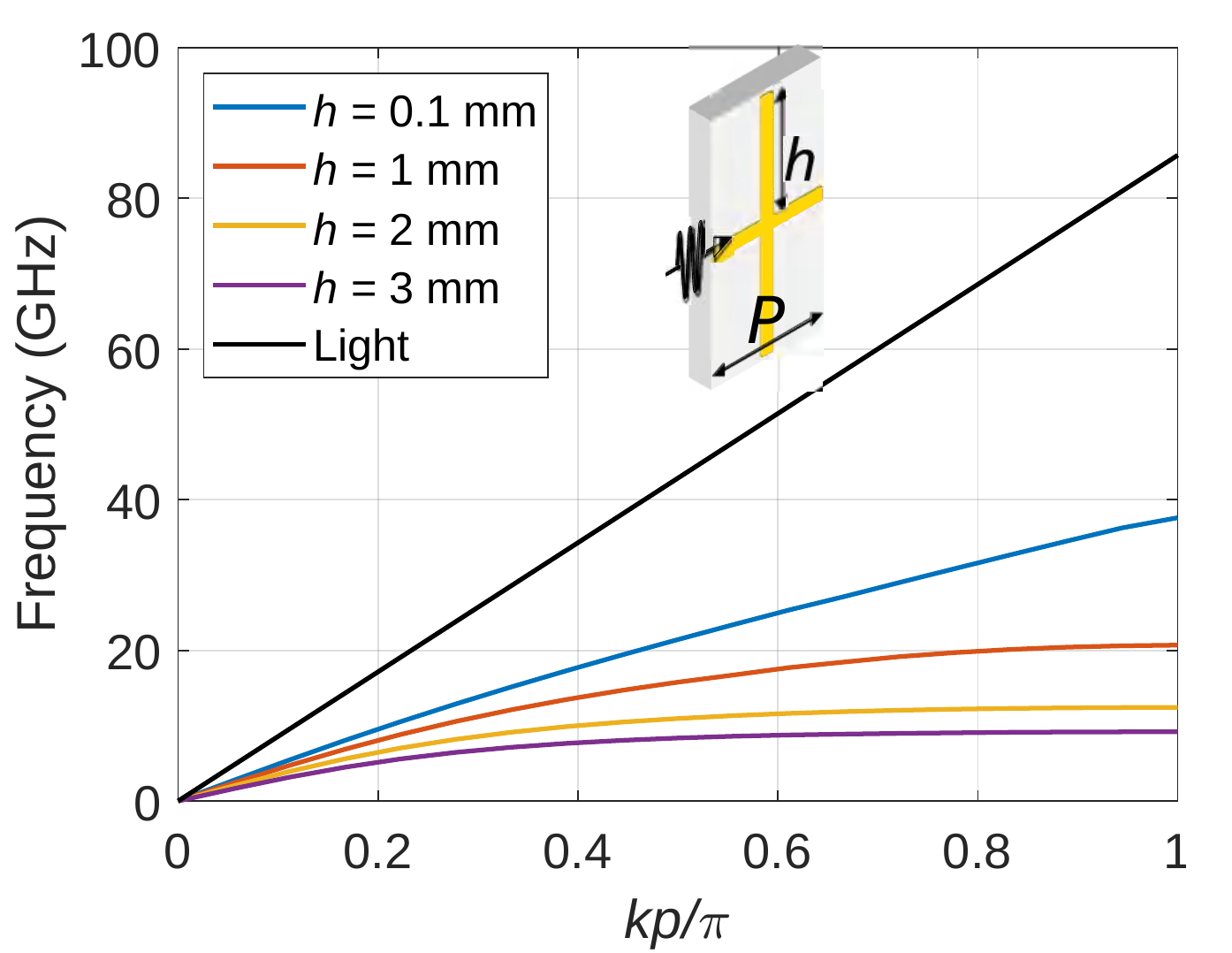}
  \caption{ Dispersion curves with different stub height $h$ of the SPPM unit cell.}
  \label{fig:dispersion curve}
\end{figure}

The fundamental mechanism for the SPPM metastructure to realize adaptive protection lies in the variation characteristics of its dispersion relation with structural dimensions. Fig. \ref{fig:dispersion curve} demonstrates the dispersion curves for different stub heights $h$ of the SPPM unit cell (shown in the inset). The polarization direction of the incident electric field is parallel to the stubs. We can find that the dispersion curves deviate from the light line.  At low frequencies, the wave number $k$ of the SPPM unit cell is close to that of the light line; as the frequency increases gradually, it becomes much larger than that of the light line until the $k$ approaches infinity—that is, the frequency approaches the asymptotic frequency. This frequency is regarded as the cut-off frequency $f_{\mathrm{c}}$ of the SPPM, endowing it with low-pass characteristics. It can also be observed that the $f_{\mathrm{c}}$ of the dispersion curve decreases gradually as the value of $h$ increases. This parameter-dependent frequency tuning capability enables flexible regulation of the wave propagation shielding threshold, laying a foundation for adaptive protection across wide frequency bands. Specifically, by incorporating NPMS devices onto the SPPM stubs, the variation of $h$ is equivalently realized through NPMS switching, thereby achieving switching between wave transmission and shielding states in the target operating frequency band.

\subsection*{Protective microstrip patch antenna}

\begin{figure}[H]
  \centering
  \includegraphics[width=0.5\linewidth]{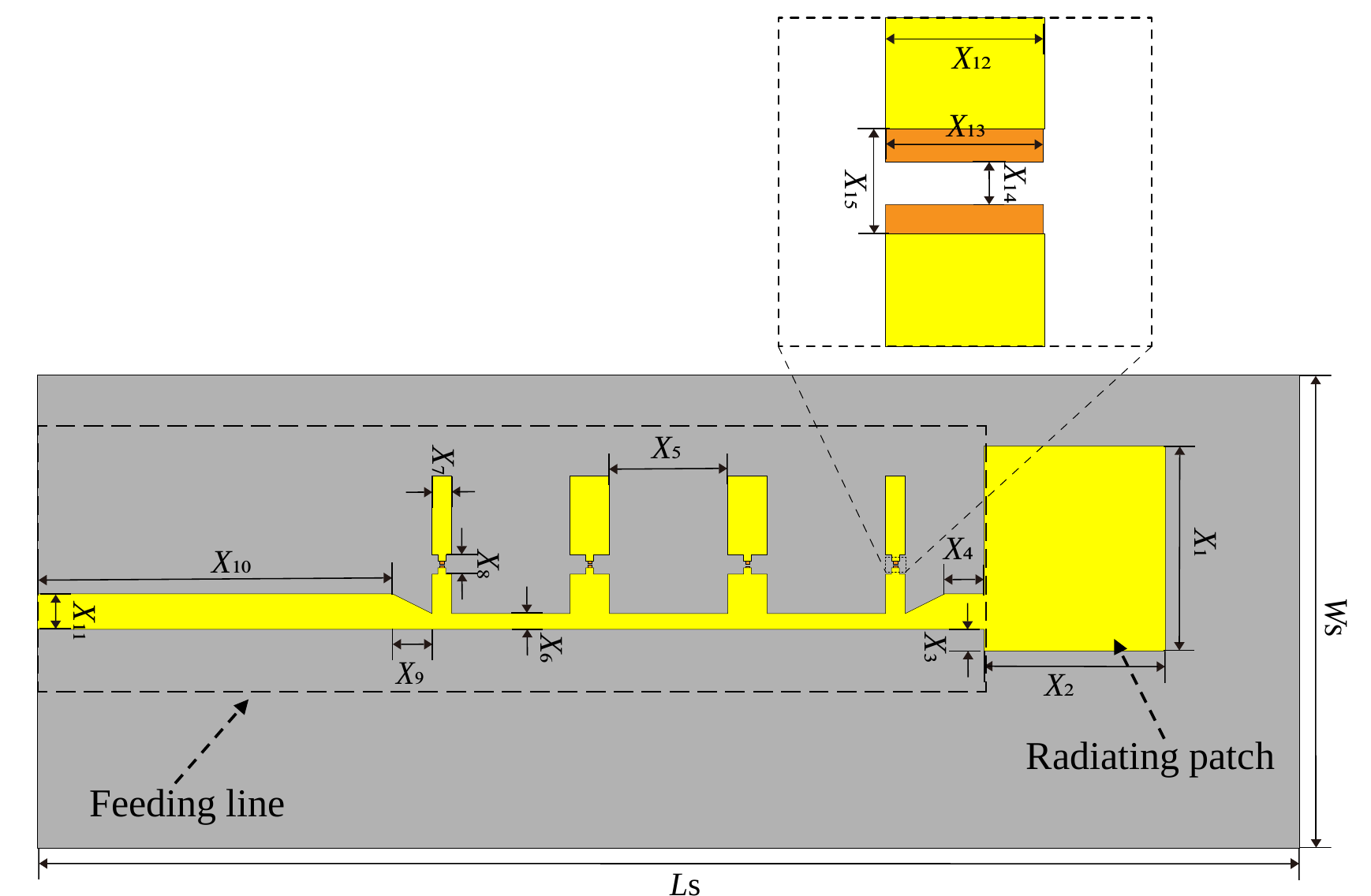}
  \caption{Structural parameters of the proposed protective microstrip patch antenna.}
  \label{fig:antenna}
\end{figure}

\begin{table}[htbp]
\centering
\caption{Detailed structural parameters of the designed protective microstrip patch antenna}
\label{tab:antenna_params}
\renewcommand{\arraystretch}{1.2}
\scriptsize
\setlength{\tabcolsep}{5pt} 
\begin{tabular}{@{}lccccccccc@{}}
\hline
\textbf{Param.} & $L_s$ & $W_s$ & $X_1$ & $X_2$ & $X_3$ & $X_4$  & $X_5$  & $X_6$  & $X_7$ \\
\textbf{Value}  & 16 mm & 6 mm & 2.6 mm & 2.3 mm & 0.275 mm& 0.5 mm &1.5 mm & 0.2 mm&0.25 mm\\
\hline
\textbf{Param.} & $X_8$ & $X_9$ & $X_{10}$ & $X_{11}$  & $X_{12}$  & $X_{13}$  & $X_{14}$ & $X_{15}$ \\
\textbf{Value}  & 0.2 mm & 0.5 mm & 4.5 mm & 0.45 mm & 0.1 mm & 0.1 mm &800 nm & 40 \si{\micro\meter}& \\
\hline
\end{tabular}
\end{table}

\subsection*{Limiter}

\begin{figure}[H]
  \centering
  \includegraphics[width=0.5\linewidth,angle=270]{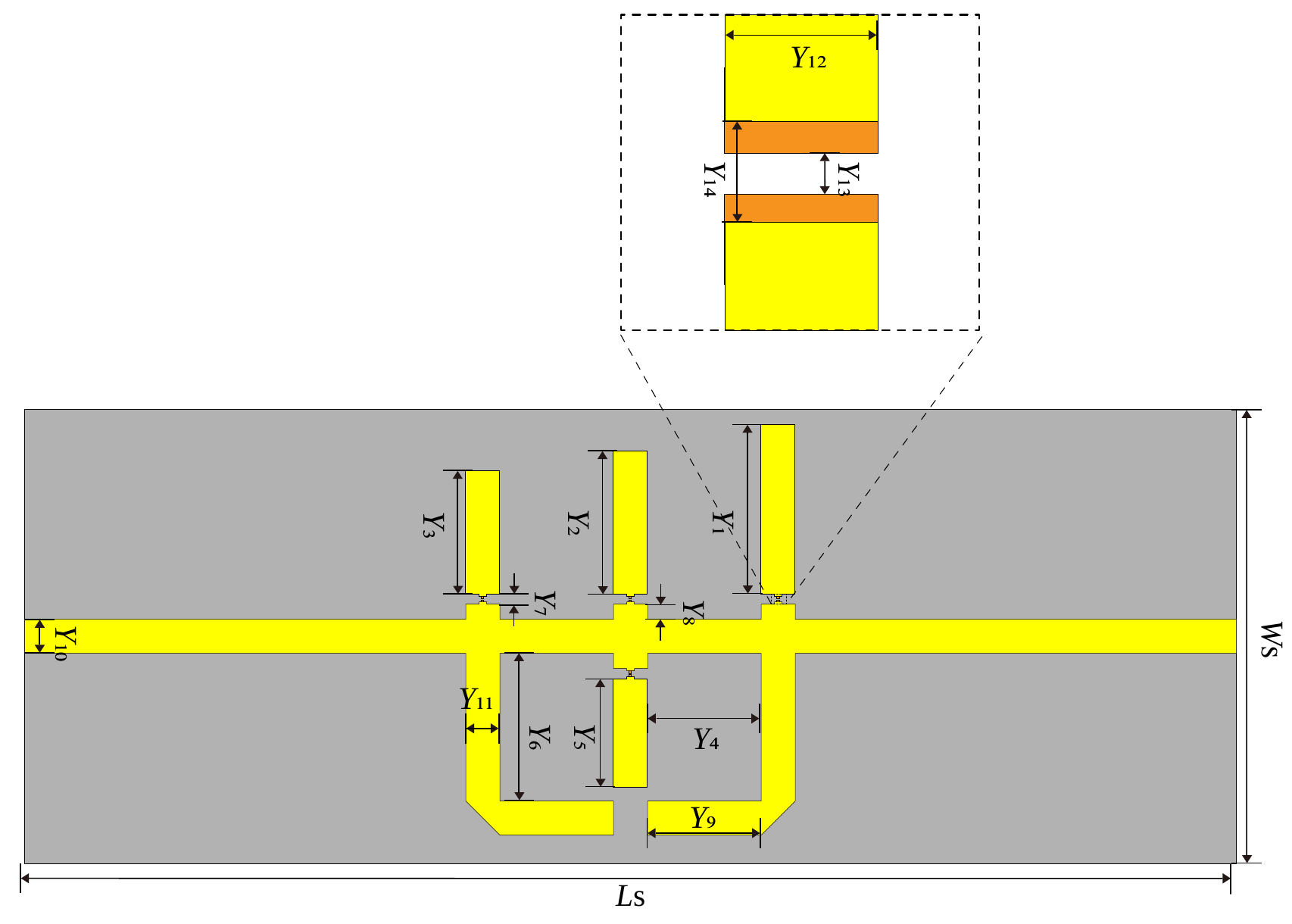}
  \caption{Structural parameters of the proposed Limiter.}
  \label{fig:limiter}
\end{figure}

\begin{table}[htbp]
\centering
\caption{Detailed structural parameters of the designed limiter}
\label{tab:circuit_params}
\renewcommand{\arraystretch}{1.2}
\scriptsize
\setlength{\tabcolsep}{5pt} 
\begin{tabular}{@{}lcccccccc@{}}
\hline
\textbf{Param.} & $L_s$ & $W_s$ & $Y_1$ & $Y_2$ & $Y_3$ & $Y_4$  & $Y_5$  & $Y_6$ \\
\textbf{Value}  & 16 mm & 6 mm & 2.27 mm & 1.92 mm & 1.65 mm & 1.5 mm &1.45 mm & 1.95 mm\\
\hline
\textbf{Param.}  & $Y_7$ & $Y_8$ & $Y_9$ & $Y_{10}$ & $Y_{11}$  & $Y_{12}$  & $Y_{13}$  & $Y_{14}$\\
\textbf{Value}  & 40 \si{\micro\meter} & 0.23 mm & 1.5 mm & 0.45 mm& 0.45 mm & 0.1 mm &800 nm &  40 \si{\micro\meter}  \\
\hline
\end{tabular}
\end{table}

\section{HPM excitation experiments}
\begin{figure}[H]
  \centering
  \includegraphics[width=0.6\linewidth,]{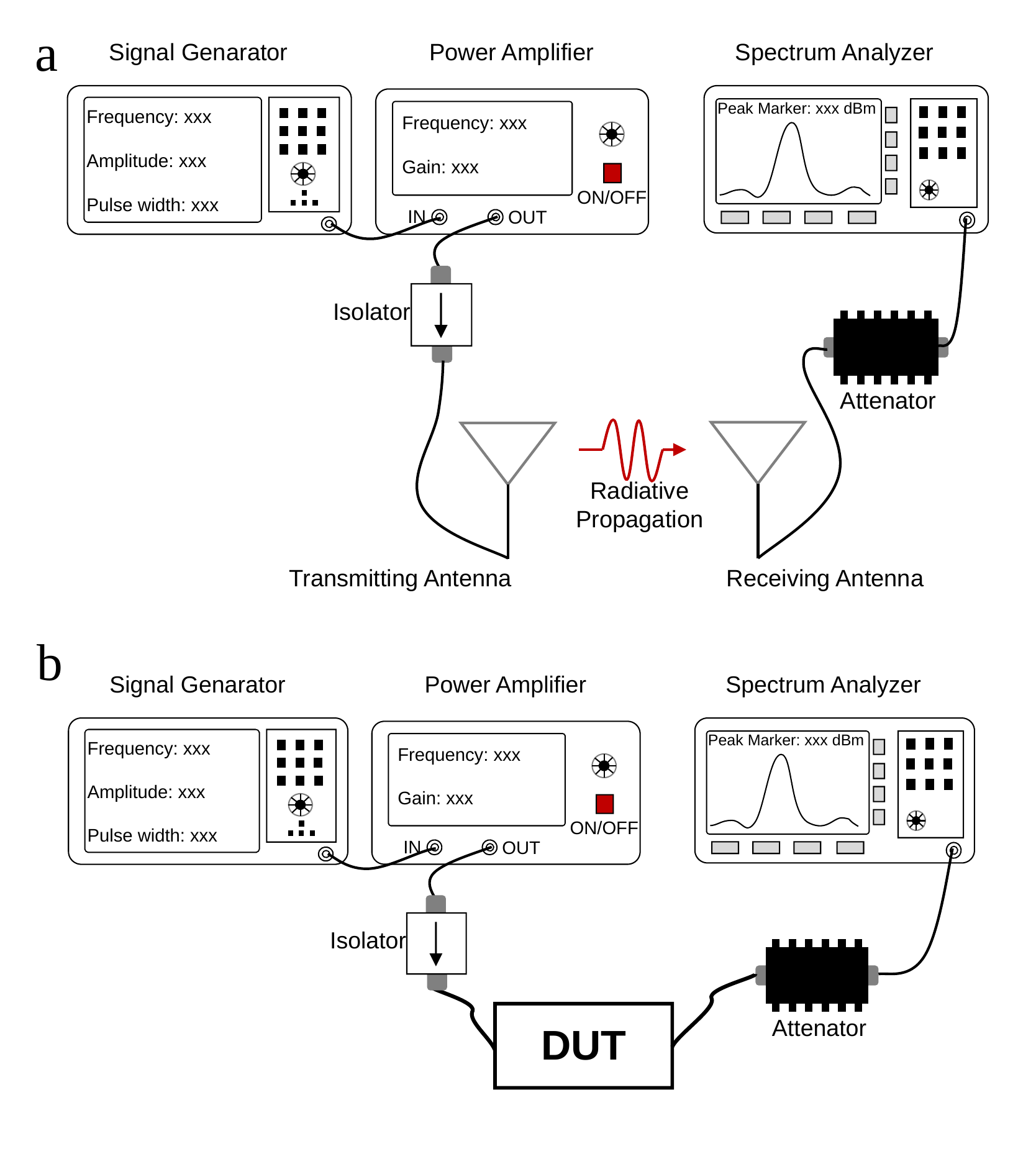}
  \caption{Schematics of experimental setups for HPM excitation. \textbf{a,} Field-level test and \textbf{b,} circuit-level test.}
  \label{fig:HPM_test_setup}
\end{figure}

Fig.~\ref{fig:HPM_test_setup} presents the schematics of the experimental setups for characterizing the responses of the proposed protective designs under HPM excitations.  A microwave signal is amplified by a power amplifier and subsequently delivered to the transmitting antenna (Fig.~\ref{fig:HPM_test_setup}a, corresponding to field-level test) or the device under test (DUT) (Fig.~\ref{fig:HPM_test_setup}b, corresponding to circuit-level test) through an isolator. The isolator is employed to suppress reflected power and protect the upstream microwave signal source and power amplifier. The received signal is passed through an attenuator before being recorded using a spectrum analyzer (or oscilloscope). For field-level tests, an extra radiative propagation link is required.

\subsection*{SPPM}
\begin{figure}[H]
  \centering
  \includegraphics[width=0.6\linewidth,]{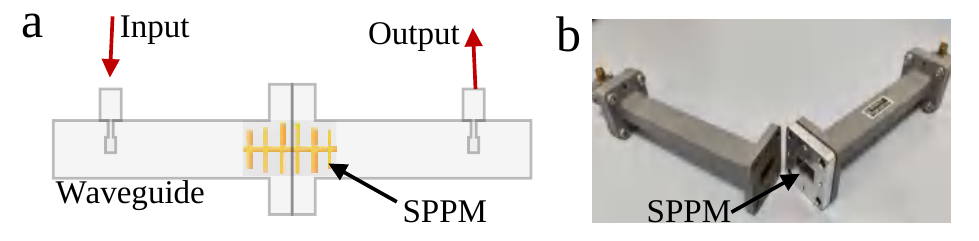}
  \caption{\textbf{a,} Schematic  and \textbf{b,} photograph of  an SPPM placed in a WR62 waveguide. }
  \label{fig:SPPM_in_waveguide}
\end{figure}

Characterization of the  proposed SPPM structure requires an HPM field environment. Because less power is required to generate the same electric-field strength in a waveguide than in free space, we employ the waveguide injection test method to characterize the response of the SPPM prototype under different field strengths. The selection of placement position takes into account the field distribution characteristics of the TE10 mode inside the waveguide, that is, the field intensity is the highest in the middle of the waveguide cross-section and decreases toward both sides. Therefore, accroding to the working frequency and structure size, we place one SPPM element at the center of the cross section of a standard WR62 waveguide to test its performance, as shown in Fig. \ref{fig:SPPM_in_waveguide}.  Then, measurements are performed on the waveguide loaded with the SPPM element, which serves as the DUT illustrated in Fig. \ref{fig:HPM_test_setup}b.

The peak electrical field strength (\(E_{\max}\)) in waveguide is calculated as follows:
\begin{equation}
E_{\max}= 2\sqrt{\frac{P\eta}{ab}}\,\frac{1}{\sqrt{1-\left(\dfrac{f_c}{f}\right)^2}},
\end{equation}
\begin{equation}
f_c=\frac{3\times10^8}{2a},\qquad \eta=\frac{377}{\sqrt{1-\left(\dfrac{f_c}{f}\right)^2}}.
\end{equation}
where \(f_c\) is the \(\mathrm{TE}_{10}\)-mode cut-off frequency of the waveguide, \(P\) is the input power, and \(a\) and \(b\) are the width and
height of the waveguide cross section, respectively. For the standard WR62 waveguide, \(a\) = 15.799 mm and  \(b\) = 7.899 mm.

\begin{table}[H]
    \centering
    \caption{Peak electric field strength (Unit: $\mathrm{kV/m}$) at different frequencies and input powers in the WR62 waveguide}
    \label{tab:wr62_field_strength}
    \renewcommand{\arraystretch}{1.5}
    \scriptsize
    \setlength{\tabcolsep}{5pt}
    \begin{tabular}{c|rrrrrr}
        \hline
        \diagbox{Freq. (GHz)}{Power (dBm)}
            & 10 & 20 & 30 & 40 & 50 & 60 \\
        \hline
        12 & 0.444 & 1.403 & 4.436 & 14.028 & 44.360 & 140.280 \\
        13 & 0.411 & 1.301 & 4.113 & 13.007 & 41.130 & 130.070 \\
        14 & 0.396 & 1.253 & 3.961 & 12.525 & 39.610 & 125.250 \\
        15 & 0.387 & 1.225 & 3.873 & 12.248 & 38.730 & 122.480 \\
        16 & 0.382 & 1.207 & 3.817 & 12.071 & 38.170 & 120.710 \\
        17 & 0.378 & 1.195 & 3.779 & 11.949 & 37.790 & 119.490 \\
        18 & 0.375 & 1.187 & 3.752 & 11.865 & 37.520 & 118.650 \\
        \hline
    \end{tabular}
\end{table}

\begin{figure}[H]
  \centering
  \includegraphics[width=\linewidth]{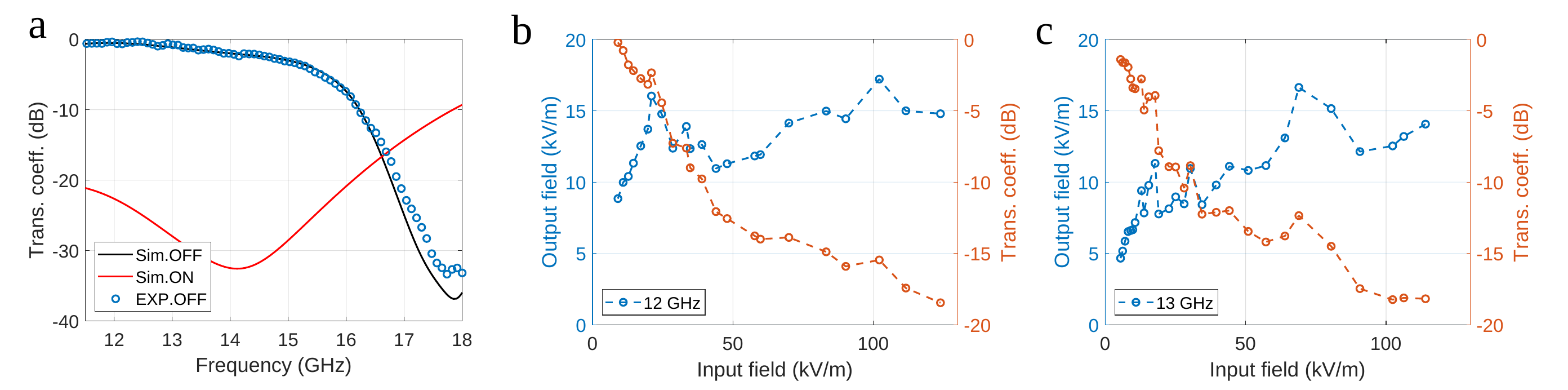}
  \caption{ Results for SPPM-2. \textbf{a,} Transmission coefficients. Nonlinear responses under HPM field excitations at \textbf{b,} 12 GHz and  \textbf{c,} 13GHz.}
  \label{fig:SSPP-2_result}
\end{figure}

The results for SPPM-2 through waveguide injection tests are presented in Fig. \ref{fig:SSPP-2_result}. The corresponding results for SPPM-1 have already been presented in the main text. As can be seen, the insertion loss is less than 1 dB below 13 GHz and less than 2 dB below 14 GHz, which is consistent with the full-wave simulation results. At 12~\si{GHz} and 13~\si{GHz}, the output fields were limited to $1.48\times10^4$~\si{V/m} and $1.41\times10^4$~\si{V/m}, respectively, corresponding to shielding effectiveness (SE) values of approximately 18.46~\si{dB} and 18.18~\si{dB} (Fig.~\ref{fig:SSPP-2_result}b,c).

\begin{figure}[H]
  \centering
  \includegraphics[width=0.5\linewidth,]{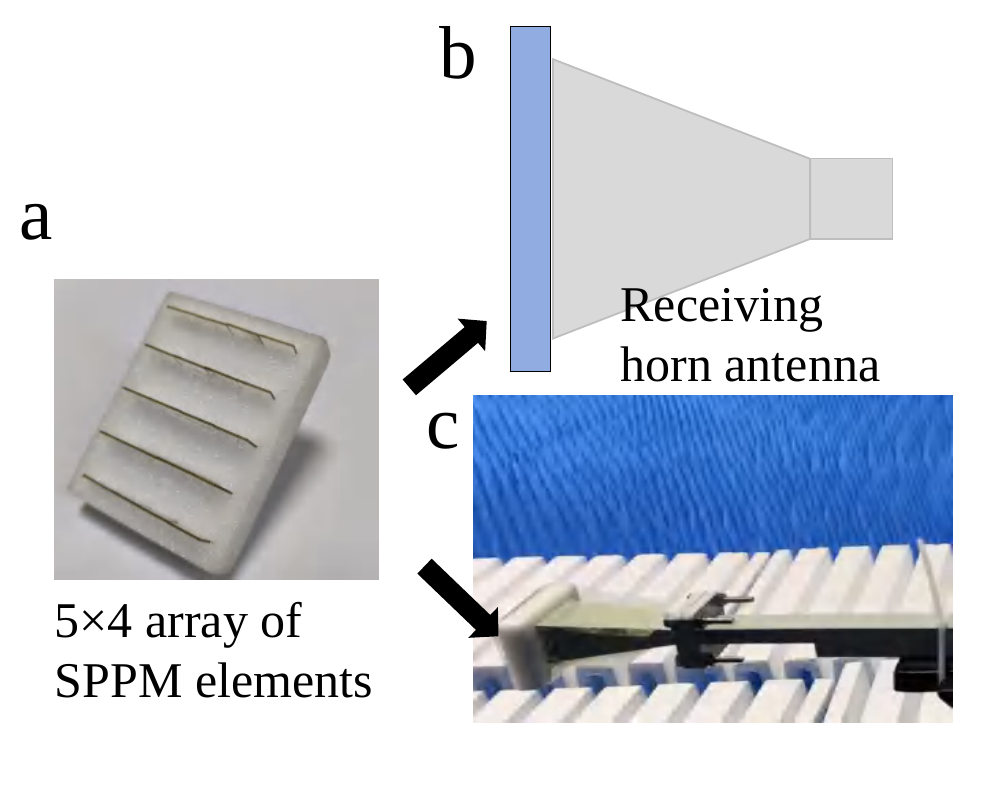}
  \caption{\textbf{a,} The $5~\times~4$ array of SPPM-2 elements placed in front of the receiving horn antenna. \textbf{b,} Placement schematic and \textbf{c,} photograph.}
  \label{fig:array}
\end{figure}

\begin{figure}[H]
  \centering
  \includegraphics[width=0.7\linewidth,]{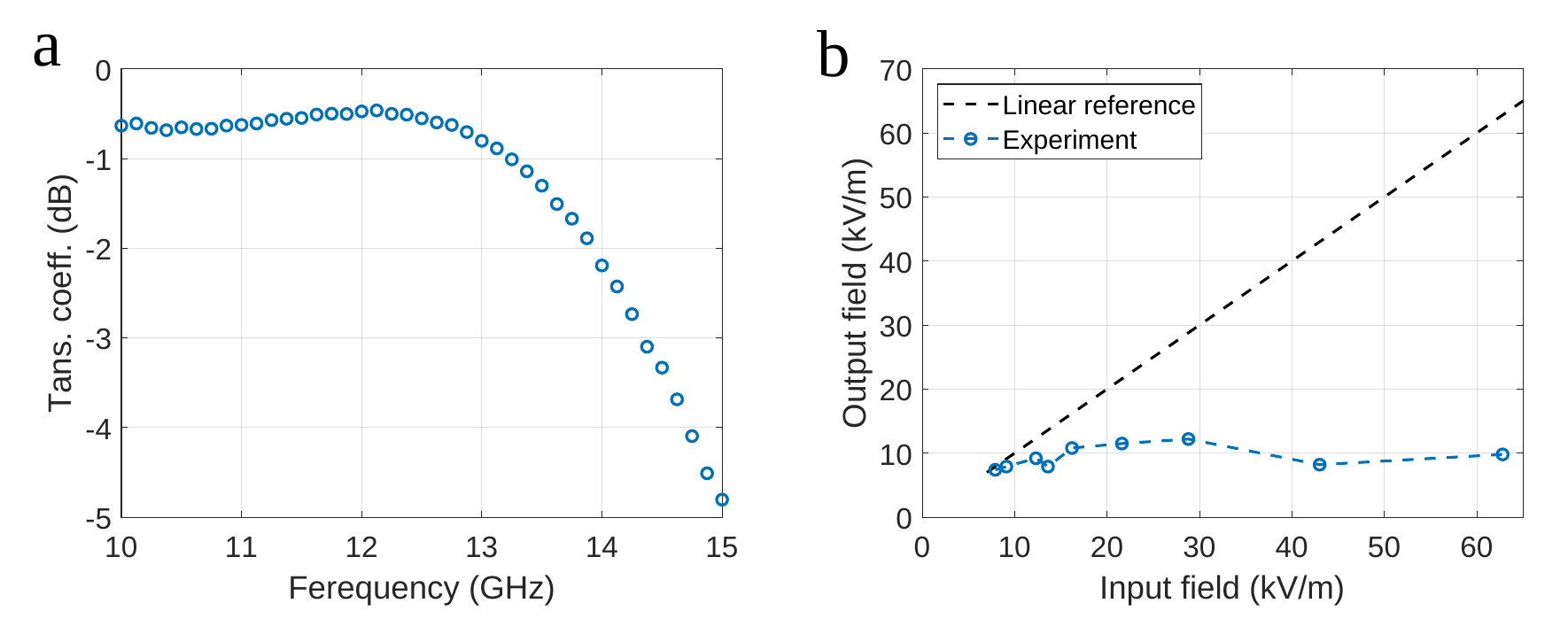}
  \caption{\textbf{a,} Insertion loss results for the SPPM array. \textbf{b,} Response under HPM field excitation at 14.25 GHz.}
  \label{fig:array_result}
\end{figure}

We then fabricate a 5 $\times$ 4 array of SPPM-2 elements (Fig. \ref{fig:array}a) and characterize its performance under spatial irradiation in a microwave anechoic chamber. The spacing between two adjacent parallel SPPM elements in the array is 10 mm. The entire array is placed in front of the horn antenna, as shown in the Fig. \ref{fig:array}. Then, the horn antenna loaded with the array serves as the receiving antenna, as illustrated in Fig. \ref{fig:HPM_test_setup}a. The results of the array are shown in Fig. \ref{fig:array_result}, and agree well with the waveguide injection test results shown in Fig. \ref{fig:SSPP-2_result}.

\subsection*{Protective microstrip patch antenna}
\begin{figure}[H]
  \centering
  \includegraphics[width=0.4\linewidth,]{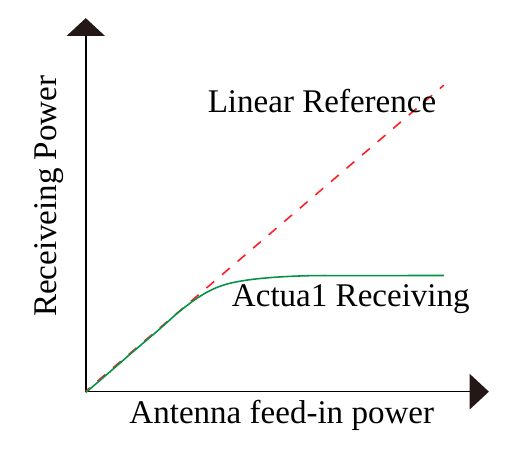}
  \caption{Nonlinear response schematic of the proposed protective antenna. }
  \label{fig:Antenna_response_curve}
\end{figure}

A dedicated high-power radiation source covering the operational frequency band of our designed antenna is unavailable; hence, a spatial field of tens of kV/m strength cannot be constructed to directly excite the designed protective microstrip patch antenna. Therefore, we employ the designed protective antenna as the transmitting antenna\cite{FangWu-308}. Electromagnetic waves radiated by the proposed antenna are captured by a standard receiving antenna to monitor variations in its radiation capability. By gradually increasing the incident microwave power and monitoring the received signal power, the transmission response of the proposed antenna is characterized under different excitation intensities. The radiation capability is then assessed from the suppression of the received signal at high incident power relative to the low-power transmission response. A decline in the radiation capability indicates that the proposed antenna enters the protection state, and accurate protection effectiveness can be calculated by comparing the radiation performance before and after activation (Fig. \ref{fig:Antenna_response_curve}). Since the antenna is a reciprocal device, the variation in receiving capability can be inferred from the changes in its radiation capability, which enables the evaluation of the protection effectiveness of the proposed antenna.

The relationship between the received power \(P_r\) at the antenna terminals and the incident time-averaged power density $S_{\mathrm{av}}$ (in W/m\(^2\)) is given by
\begin{equation}
P_r = S_{\mathrm{av}} \, A_e
\label{eq:antenna1}
\end{equation}
where \(A_e\) is the effective aperture of the antenna. For a lossless antenna with gain \(G\), the effective aperture is
\begin{equation}
A_e = \frac{G \lambda^2}{4\pi}
\label{eq:antenna2}
\end{equation}
with \(\lambda = c/f\) being the free‐space wavelength.

The power density of a plane wave is related to the peak electric field strength $ E_{\mathrm{peak}}$ (in V/m) by
\begin{equation}
S_{\mathrm{av}} = \frac{E_{\mathrm{peak}}^2}{2\eta_0}
\label{eq:antenna3}
\end{equation}
where \(\eta_0 \approx 120\pi \approx 376.73\,\Omega\) is the intrinsic impedance of free space.

Combining Eq. \eqref{eq:antenna1} -- \eqref{eq:antenna3}, we obtain the direct relation
\begin{equation}
E_{\mathrm{peak}} = \sqrt{\frac{8\pi \, P_r \, \eta_0}{G \, \lambda^2}}
\label{eq:final}
\end{equation}

For example, for an antenna with a gain of 3.0 dBi operating at 17.5 GHz, a received power of 56 dBm corresponds to an incident peak electric field strength $ E_{\mathrm{peak}}$ of about 80.2 kV/m. 
In our measurement of the proposed protective antenna, which has a measured gain of 3.0 dBi, the antenna serves as the transmitting antenna.  Based on antenna reciprocity, when the power of 56 dBm is fed into the protective antenna, the excitation borne by the NPMS devices  is equivalent to the antenna being subjected to an incident peak field strength $ E_{\mathrm{peak}}$ of 80.2 kV/m.


\textbf{Normalized power density figure-of-merit (FOM)}: Let \(A_{\mathrm{phys}}\) denote the physical aperture area,
\(A_{\mathrm{e}}\) the effective aperture (Eq. \eqref{eq:antenna2}), \(S_{\mathrm{av}}\) the incident time-averaged flux density  (Eq. \eqref{eq:antenna3}), and \(\eta_{\mathrm{ap}}\) the aperture efficiency.
The aperture efficiency is defined as
\begin{equation}
    \eta_{\mathrm{ap}}
    =
    \frac{A_{\mathrm{e}}}{A_{\mathrm{phys}}}.
    \label{eq:aperture_efficiency}
\end{equation}

It follows from equation\eqref{eq:aperture_efficiency} that
\begin{equation}
    A_{\mathrm{e}}
    =
    \eta_{\mathrm{ap}} A_{\mathrm{phys}}.
    \label{eq:effective_aperture}
\end{equation}

For an incident plane wave, the received power is
\begin{equation}
    P_{\mathrm{r}}
    =
    S_{\mathrm{av}} A_{\mathrm{e}}
    =
    S_{\mathrm{av}} \eta_{\mathrm{ap}} A_{\mathrm{phys}}.
    \label{eq:received_power}
\end{equation}

The normalized power density figure-of-merit (FOM) is defined as
\begin{equation}
    \mathrm{FOM}
    =
    \frac{P_{\mathrm{r}}\eta_{\mathrm{ap}}}
         {A_{\mathrm{phys}}}.
    \label{eq:metric_definition}
\end{equation}

Substituting Eq. \eqref{eq:received_power} into Eq. \eqref{eq:metric_definition} gives
\begin{equation}
    \begin{aligned}
        \mathrm{FOM}
        &=
        \frac{
            \left(
                S_{\mathrm{av}}\eta_{\mathrm{ap}}A_{\mathrm{phys}}
            \right)\eta_{\mathrm{ap}}
        }{
            A_{\mathrm{phys}}
        } \\
        &=
        S_{\mathrm{av}}\eta_{\mathrm{ap}}^{2}=\frac{E_{\mathrm{peak}}^2}{2\eta_0}\eta_{\mathrm{ap}}^{2}.
    \end{aligned}
    \label{eq:metric_result}
\end{equation}

For a fixed incident power flux density, the  normalized power density FOM is proportional to the square of the aperture efficiency. The physical aperture area is canceled because the received power is proportional to that area.

\begin{table}
\centering
\caption{Size of diode-based protective antennas and the proposed protective antenna.}
\label{tab:Antenna_Size}
\renewcommand{\arraystretch}{2}
\scriptsize
\setlength{\tabcolsep}{5pt} 
\begin{tabular}{@{}lcccccccc@{}}
\hline
\textbf{Ref. } &\cite{DengLin-375} &\cite{FangWu-376}  &\cite{Zhasong} &  \cite{QuZha-378} & \cite{QuZha-379} &\cite{WangTang-380}   &  \cite{ZhaLiu-374} & This work     \\
\hline
\textbf{Size. (\si{mm^2})}  &$35\times50$ &$144\times60$  &$200^2\times\pi$&$250^2\times\pi$   & $180^2$  & $26\times30$ &$250^2\times\pi$ &$14\times2.6$ \\
\hline
\end{tabular}
\end{table}

\section{Effect of flexible substrate bending on the electrode gap}

\begin{figure}[H]
  \centering
  \includegraphics[width=0.5\linewidth,]{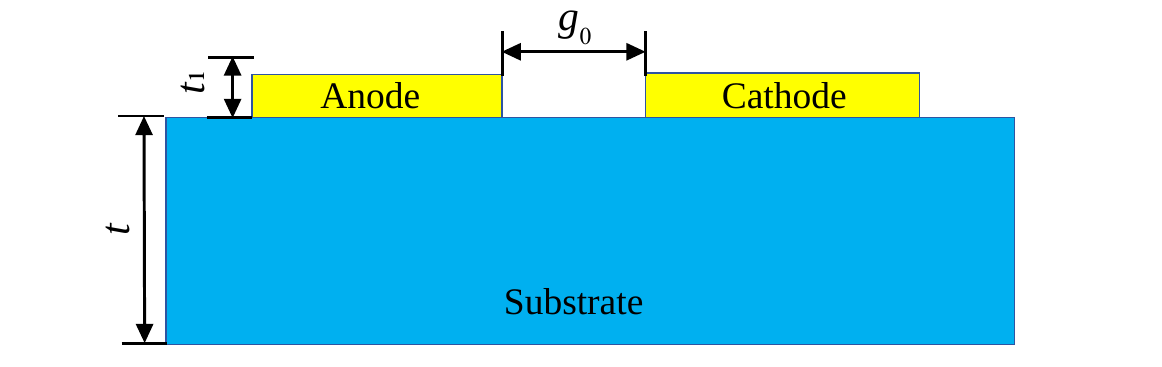}
  \caption{Diagram of parallel electrodes fabricated on a flexible substrate.}
  \label{fig:flexible}
\end{figure}

Consider two parallel electrodes fabricated on a flexible substrate (Fig. \ref{fig:flexible}). Here the case in which the bending axis is parallel to the electrode edges is considered. In this configuration, the bending-induced tensile or compressive strain acts directly along the gap direction.

\begin{figure}[H]
  \centering
  \includegraphics[width=0.6\linewidth,]{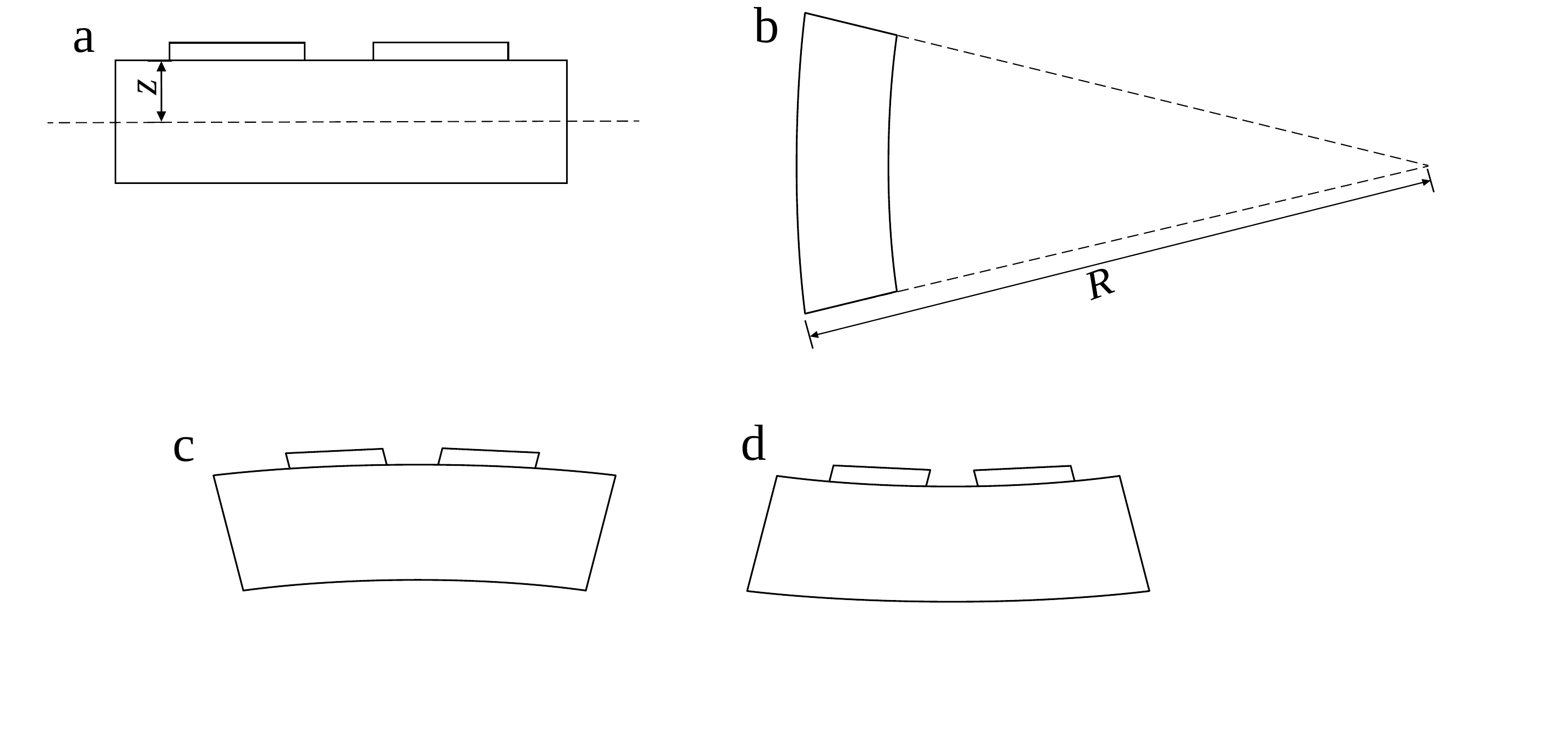}
  \caption{Diagram of a flexible substrate under bending. \textbf{a,} Without bending. \textbf{b,} Bending radius. \textbf{c,}Tensile bending. \textbf{d,} Compressive bending.}
  \label{fig:flexible_bending}
\end{figure}

Let the initial electrode gap be \(g_0\), the distance from the neutral mechanical plane to the electrode surface be \(z\), and the bending radius be \(R\) (Fig. \ref{fig:flexible_bending}a,b). Under small-deformation pure bending, the strain at the electrode surface is approximately\cite{2017Mechanics,Hwanhee}

\begin{equation}
\varepsilon_{\mathrm{gap}} \approx \frac{z}{R},
\end{equation}

where the sign of \(z/R\) depends on whether the electrode surface is under tension (Fig. \ref{fig:flexible_bending}c) or compression (Fig. \ref{fig:flexible_bending}d). The deformed gap width is therefore

\begin{equation}
g \approx g_0\left(1+\varepsilon_{\mathrm{gap}}\right)
  = g_0\left(1+\frac{z}{R}\right).
\end{equation}

Accordingly, the change in the gap width is

\begin{equation}
\Delta g = g-g_0 \approx g_0\frac{z}{R}.
\end{equation}

When the electrodes are located on the outer, convex side of the bent substrate, the electrode surface is under tensile strain. Thus,

\begin{equation}
\varepsilon_{\mathrm{gap}}>0,
\qquad
g>g_0,
\end{equation}

and the electrode gap increases.

When the electrodes are located on the inner, concave side of the bent substrate, the electrode surface is under compressive strain. Thus,

\begin{equation}
\varepsilon_{\mathrm{gap}}<0,
\qquad
g<g_0,
\end{equation}

and the electrode gap decreases.

For a single-layer substrate of thickness \(t\), if the neutral plane is approximately located at the mid-plane of the substrate and the electrode thickness is negligible, then

\begin{equation}
z \approx \frac{t}{2}.
\end{equation}

The magnitude of the gap change can consequently be estimated as

\begin{equation}
\left|\Delta g\right|
\approx
g_0\frac{t}{2R}.
\end{equation}

For example, consider

\begin{equation}
t=20~\si{\micro\meter},
\qquad
R=1~\mathrm{mm},
\qquad
g_0=100~\si{\nano\meter}.
\end{equation}

The strain magnitude at the electrode surface is

\begin{equation}
\left|\varepsilon_{\mathrm{gap}}\right|
\approx
\frac{10~\si{\micro\meter}}{1~\mathrm{mm}}
=0.01,
\end{equation}

which corresponds to \(1\%\). The magnitude of the gap variation is therefore

\begin{equation}
\left|\Delta g\right|  \approx
100~\si{\nano\meter}\times 0.01  
=
1~\si{\nano\meter}.
\end{equation}

Hence, the gap becomes approximately \(101~\si{\nano\meter}\) under tensile bending and \(99~\si{\nano\meter}\) under compressive bending. Although a bending strain of 1\% is generally regarded as a significant deformation\cite{Qin}, the resulting geometric change in the electrode gap of the NPMS on the substrate is extremely small, leading to a negligible impact on its electrical performance. Therefore, this switch presents a highly attractive prospect for integration into flexible electronic applications, such as wearable devices and flexible sensors.


\bibliographystyle{sn-nature}
\bibliography{Bibliography}